\documentclass[a4paper,authoryear,3p,twocolumn]{elsarticle}
\pdfoutput=1

\usepackage{color}
\definecolor{linkblue}{rgb}{0,0,.8}
\definecolor{linkgreen}{rgb}{0,0.45,0}
\definecolor{urlblue}{rgb}{0,0,0.9}
\definecolor{purple}{rgb}{0.7,0.0,0.4}
\usepackage[colorlinks,bookmarks]{hyperref}
\AtBeginDocument{
\hypersetup{linkcolor=linkblue,
            citecolor=purple,
            urlcolor=urlblue}}

\usepackage[T1]{fontenc}
\usepackage[utf8]{inputenc}
\usepackage{lmodern}
\usepackage[english]{babel}
\usepackage{amsmath, amssymb}

\usepackage{array}

\newcolumntype{N}{@{}m{0pt}@{}}

\addto\captionsenglish{}

\newcommand{\dd}{\textrm{d}}
\newcommand{\betao}{\boldsymbol{\beta_{\rm o}}}
\newcommand{\betasn}{\boldsymbol{\beta_{\rm SN}}}
\newcommand{\nhat}{\boldsymbol{\hat{n}}}
\newcommand{\vhat}{\boldsymbol{\hat{\beta}_{\rm o}}}

\def \apjs {ApJS}
\def \apjl {ApJL}
\def \aap {A\&A}

\begin{document}
\begin{frontmatter}

\title{Turning noise into signal: learning from the scatter in the Hubble diagram}

\author{Tiago Castro}

\author{Miguel Quartin}

\author{Sandra Benitez-Herrera}
\address{Instituto de Física -- Universidade Federal do Rio de Janeiro, 21941-972, Rio de Janeiro, RJ, Brazil}




\begin{abstract}
    The supernova (SN) Hubble diagram residual contains valuable information on both the present matter power spectrum and its growth history. In this paper we show that this information can be retrieved with precision by combining both peculiar velocity and weak-lensing analysis on the data. To wit, peculiar velocity induces correlations on the nearby SN while lensing induces a non-Gaussian dispersion in faraway objects. We show that both effects have almost orthogonal degeneracies and discuss how they can be extracted simultaneously from the data. We analyze the JLA supernova catalog in a 14-dimensional parameter space, assuming a flexible growth-rate index $\gamma$. We arrive at the following marginalized constraints: $\sigma_8 = 0.65^{+0.23}_{-0.37}$ and $\gamma = 1.38^{+1.7}_{-0.65}$. Assuming instead GR as the correct gravitation theory (and thus $\gamma \equiv 0.55$), the constraints in $\sigma_8$ tighten further: $\sigma_8 = 0.40^{+0.21}_{-0.23}$. We show that these constraints complement well the ones obtained from other datasets and that they could improve substantially with more SNe.
\end{abstract}


\begin{keyword}
    peculiar velocity -- gravitational lensing: weak -- cosmology: observations -- cosmological parameters --  large-scale structure of the universe  -- stars: supernovae: general
\end{keyword}

\end{frontmatter}

\section{Introduction}\label{sec:intro}

Type Ia supernovae (SNe) are still the only established high-redshift standard candles, with an intrinsic dispersion $\sigma_{\rm int}$ of less than $0.15$ in magnitude after standardization~\citep{hamuy1996,riess1996}. In the late 90s, using a sample of about 50 of these objects the discovery of the accelerated expansion was established by~\cite{riess1998, perlmutter1999}. The confidence in that result was enhanced soon after in~\cite{Bahcall:1999xn} by combining the SNe with galaxy surveys early Cosmic Microwave Background (CMB) data on large-scales. It is nowadays a well-established result, with strong evidence coming from observations of the Baryonic Acoustic Oscillations~\citep{eisenstein2005, blake2011} and of the anisotropies of the CMB~\citep{Ade:2015xua}. 

Since those pioneering works, a vast number of SNe surveys have been conducted, and more are underway or planned. This will increase the number of observed explosions from the current $\sim10^{3}$~\citep{Betoule:2014frx} to over $10^{6}$~\citep{Abell:2009aa}.
However, although the statistical uncertainty associated with the intrinsic dispersion in magnitude will consequently be improved by more than one order of magnitude, it will be a daunting task to improve the systematics at the same rate. These are therefore likely to become dominant in the upcoming large datasets.

Interestingly, this increasing contamination from systematics opens up new opportunities. In particular, a good part of this extra ``noise'' in the Hubble diagram can be converted into ``signal'' by modeling two independent astrophysical effects: peculiar velocities (PV) and gravitational lensing. The former introduces correlations in the supernova magnitudes for $z \lesssim 0.1$~\citep{gordon2007}. The latter introduces a redshift-dependent non-Gaussian scatter in the distribution of the supernovae and was first discussed in~\cite{Bernardeau:1996un,Hamana:1999rk,Valageas:1999ir}. Contrary to peculiar velocities, lensing introduces \emph{no} correlations among the supernovae,\footnote{Unless more than one SNe have really small angular separation and happen to be lensed by the same structure.} and is moreover only relevant for high redshift events ($z \gtrsim 0.4$). These new sources of signal allow the SNe to transcend their background role and constrain cosmological perturbation parameters.

In the case of lensing, this idea was proposed a decade ago by~\cite{Dodelson:2005zt}, and recently developed further in~\cite{Quartin:2013moa, Castro:2014oja} in what was called the Method of the Moments (MeMo). The idea behind it is rather simple. The apparent magnitude of standard candles depends on the gravitational perturbations crossed by their light along its geodesic. The geodesic is deflected by any intervening matter, but the effect is in practice dominated by the distribution of matter on the largest scales. Thus, if we analyze the statistical distribution of the SNe apparent magnitudes around their mean value, we can constrain properties of the intervening  matter, like  the amplitude of the linear perturbations or parameters of the halo model.

The weak-lensing effect can be encapsulated in the convergence lensing PDF, which gives the probability density of having a given convergence (and thus a given magnification, as to first order the shear can be neglected) as a function of redshift. This PDF was shown to have an approximate log-normal shape in~\cite{Holz:2004xx,Amendola:2013twa} (see~\ref{sec:mock-app} for one caveat though), the large skewness of which related to the fact that relatively few objects are behind clusters and thus get magnified: the large majority of geodesics come instead through cosmic voids, and thus get demagnified -- although see~\cite{Bolejko:2012uj}. It can be computed using ray-tracing techniques on  N-body simulations. However, this method is computationally too expensive to carry out likelihood analysis, so~\cite{Kainulainen:2009dw,Kainulainen:2010at} developed a fast stochastic method called sGL to compute the lensing PDF. The sGL relies on prescriptions for the mass
function~\citep{Jenkins:2000bv,Sheth:1999mn,Courtin:2010gx} and for the concentration parameter~\citep{Zhao:2008wd} as a function of cosmology. It yields results which are consistent both with the dark matter N-body simulations of~\cite{Hilbert:2007jd,Takahashi:2011qd}, with the theoretical calculations of~\cite{BenDayan:2013gc} and with the experimental data of~\cite{Jonsson:2009jp,Jonsson:2010wx}.

The lensing gets convolved with the intrinsic SNe distribution~\citep{Amendola:2010ub,Quartin:2013moa}, which dilutes the non-Gaussianity. The MeMo consists in parametrizing the lensing PDF by the first moments of the distribution, which can then be propagated into the moments of the final convolved PDF and confronted with data. \cite{Quartin:2013moa} showed that this could be used to constrain $\sigma_8$ to the percent level with $10^6$ SNe. In~\cite{Amendola:2014yca} it was shown that SNe lensing can also be used to constrain modified gravity models through its effect on the growth-rate.

In practice, in order to model the lensing-induced scatter some assumptions on the intrinsic scatter of SNe are required; in particular, we assume that it is not strongly dependent on redshift. In fact, a constant scatter was shown to be currently a reasonable hypothesis using Bayesian analysis~\citep{Castro:2014oja} and it also agrees with the underlying motivation for using SNe as standard candles whose properties do not depend on distance.

The effect of peculiar velocities on the supernova is altogether different, and was developed in detail by~\cite{hui2006,davis2011}. It starts from the fact that the Hubble diagram should in principle be built using the cosmological redshift, that is, the one due to the expansion of the universe. In practice this is not an easy task as the measured redshifts are affected by non-cosmological effects arising from the  gravitational potential at the source and around us, and by the   peculiar motions of the standard candles and of the observer. The combination of cosmological $z_{\rm cosmo}$, peculiar velocity $z_{\rm pec}$ and gravitational $z_{\rm grav}$ redshifts gives the observed redshift $z_{\rm obs}$:
\begin{equation}
    (1+z_{\rm obs})=(1+z_{\rm cosmo})(1+z_{\rm pec})(1+z_{\rm grav})\,.
    \label{eq:redshift}
\end{equation}
The last two terms are systematic effects that need to be accounted for. In terms of magnitudes, a bias $\delta z$ corresponds to a bias $\delta m = \delta z \,\dd m/\dd z$. The nature of the Hubble diagram makes these corrections in magnitude large for low redshifts, where the slope is very steep.

The correction for our own velocity is straightforward: assuming our velocity to be the only contribution to the CMB dipole, Planck infers the following (curiously mnemonic) value: $\beta_{\rm o} =(1.2345 \pm 0.0007) \times 10^{-3}$~\cite{Adam:2015vua}. This assumption is theoretically well motivated (we only expect the non-kinetic dipole to be ${\cal O}(10^{-5})$ like the other multipoles), and in any case it is not an untested one: our velocity also induces an aberration in the sky. This was shown to be detectable by Planck in~\cite{Kosowsky:2010jm,Amendola:2010ty}, and consequently measured in the data by~\cite{Aghanim:2013suk}, although with only $36\%$ precision (considering systematics). \cite{Notari:2011sb} showed that this error bar could be reduced to less than $10\%$ in future CMB experiments.
\cite{Yoon:2015lta} also derived that a $20\%$ precision might be possible by measuring the kinematic dipole with future galaxy surveys.

The peculiar velocities of the SNe themselves are harder to account for. They are often just modeled as Gaussian random terms, which are included in the covariance matrix calculation. A random velocity $v=300$~km/s corresponds to an uncertainty $\sigma^{\rm pec}_{\rm m}=0.2$~mag for objects at redshift $z=0.01$, but only $\sigma^{\rm pec}_{\rm m}=0.02$~mag at $z=0.1$~\citep{davis2011}. The SNe velocities are however not really random, as the large-scale gravitational potential wells incur in coherent velocity flows, which extend to many dozens of Mpc~\citep{Hoffman:2015waa}. As discussed in~\cite{hui2006}, any two SNe in the same region of the sky will have therefore correlated magnitude fluctuations. In a nutshell, if a given supernovae is dimmer than the average because it is moving away from us, a nearby supernova has an \emph{excess probability} of also being dimmer than average because it will be in the same velocity flow. Moreover, very low-$z$ galaxies should have peculiar velocities correlated to
our own Milky Way's. These corrections should in principle not be ignored, lest we introduce biases to the inferred cosmological parameters~\citep{Neill:2007fh,gordon2007}. Some supernovae catalogs try to avoid this issue simply by not including supernovae below redshift $\sim 0.01$. As we discuss below, we can (and should) easily subtract from the data the effects from our peculiar velocity using the measurements of the CMB dipole. This should remove this possible bias, even for SNe with redshift below $0.01$. 

Importantly, the amplitude of the above correlations depend on cosmology, and it was realized in~\cite{Bonvin:2005ps} that this would open a new avenue of research. As these correlations are directly related to the 2-point correlation function of matter and, therefore, to the matter power spectrum, it was soon realized in \cite{gordon2007} that even a 271 supernovae catalog available at the time was able to put the impressive constraint $\sigma_8 = 0.79\pm0.22$. A similar idea was proposed also in~\cite{Hannestad:2007fb}. \cite{Abate:2008au} further developed the idea to show that it could be used to measure also the growth of perturbations.

Both lensing and PV are thus new sources of signal, and each allows the SNe to transcend their background role and constrain cosmological perturbation parameters. In this paper we develop a simple technique to combine them together and show that these two probes are very complementary. We develop the technique of using PV in detail and adapt the Joint Lightcurve Analysis (JLA) supernova catalog~\citep{Betoule:2014frx} in order to do so. We show that combining together PV and lensing allows one to break the degeneracy between $\sigma_8$ (which is directly related to the the amplitude of the power spectrum, as per Eq.~\eqref{eq:s8} below) and the index of growth-rate of matter perturbations $\gamma$.

This paper is organized as follows: in Section~\ref{sec:pecvel} we review the peculiar velocity effects in the Hubble diagram; in Section~\ref{sec:jla} we review the JLA catalog and adapt it for measuring the PV; in Section~\ref{sec:combined} we develop a simple technique to combine PV and lensing effects in supernova analysis; in Section~\ref{sec:sigma} we show the derived constraints on the power spectrum parameters and compare the results with the ones obtained from other datasets; finally, we discuss the conclusions and perspectives in Section~\ref{sec:conc}. We also provide some technical details on our Monte Carlo Markov chain analysis in~\ref{sec:MCMC}, on our search for systematics in the data in~\ref{sec:tests} and on our mock catalog tests in~\ref{sec:mock-app}.

\section{The effect of peculiar velocities on the Hubble diagram}\label{sec:pecvel}

\subsection{The effect on distances and magnitudes}\label{sec:pv-dist-mag}

In terms of the peculiar velocities, Eq.~\eqref{eq:redshift} can be approximated as~\citep{hui2006,davis2011}
\begin{equation}
    (1+z_{\rm obs}) = (1+z_{\rm cosmo})\big(1+\betasn \cdot \nhat - \betao \cdot \nhat\big) \,,
    \label{eq:redshift_2}
\end{equation}
where $\betasn$ and $\betao$ are the peculiar velocities (in units of the speed of light) of the supernovae and of our own peculiar motion, respectively; $\nhat$ is the unit vector for the direction between the observer and SNe. Peculiar velocities affect not only $z_{\rm obs}$, but as shown by~\cite{hui2006} also directly the luminosity of the sources due to the aberration effect, which changes the angular size of the objects much like gravitational lensing. The observed luminosity distance $d_L$ is thus related to the cosmological distance $\bar{d_{\rm L}}$ by:
\begin{equation}
    d_{\rm L}(z_{\rm obs}) = \bar{d_{\rm L}}(z_{\rm cosmo})\big(1+2\betasn \cdot \nhat - \betao \cdot \nhat \big) \,,
    \label{eq:dl_1}
\end{equation}
with $\bar{d_{\rm L}}$ given by:
\begin{equation}
    \bar{d_{\rm L}}(z) \,=\, (1+z)\int_0^z{\frac{c}{H(z')}\,\dd z'}\,.
    \label{eq:dl_1}
\end{equation}

The quantity of interest is the change $\delta_{d_{\rm L}}(z)$ in $d_{\rm L}$ at the observed redshift:
\begin{equation}
    \delta_{d_{\rm L}}(z_{\rm obs}) = \frac{[d_{\rm L}(z_{\rm obs}) - \bar{d}_{\rm L}(z_{\rm obs})]}{\bar{d}_{\rm L}(z_{\rm obs})}.
\end{equation}
If we Taylor expand  $\bar{d_{\rm L}}$, we can rewrite the above equation as
\begin{equation}
    \delta_{d_{\rm L}}(z_{\rm obs}) = \betasn\cdot \nhat  -
    \frac{(1+z_{\rm obs})^2 (\betasn - \betao)}{H(z_{\rm obs})d_{\rm L}(z_{\rm obs})}\cdot \nhat\,.
\end{equation}
Using the relation between magnitudes and luminosity distances
\begin{equation}
 	m=5\log_{10}\left(\frac{d_L}{10 \rm{Mpc}}\right)+M,
\end{equation}
and the equations for $d_L$ and $\bar{d}_L$ we can both correct the JLA magnitudes and remove the dipole term ($\betao$) from $\bar{d}_L$. The corresponding change in magnitude is
\begin{equation}
    \delta_{m}=\left(1+\frac{1+z_{\rm obs}}{H \chi}\right)(\betasn \cdot \nhat - \betao \cdot \nhat)\,.
\end{equation}

Finally, aberration also causes the change in apparent position of the sources in the sky, and thus the inferred distance between them. This effect is not usually discussed in the literature, and to our knowledge has been so far ignored, possibly on the grounds that in practice the overall effect is small. We nevertheless discuss this in Section~\ref{sec:jla}.

\subsection{Peculiar Velocity covariance}\label{sec:pv-covar}

From now on we will denote the observed redshift $z_{\rm obs}$ as just $z$ for simplicity of notation.

The velocity correlation function is defined by
\begin{equation}
    \xi ({\bf r_i},{\bf{r_j}}) \equiv \big\langle ({\bf \betasn(r_i) \cdot \hat{r}_i})({\bf \betasn(r_j) \cdot \hat{r}_j})\big\rangle\,,
\end{equation}
where ${\bf \hat{r}_i}$, ${\bf \hat{r}_j}$ are the unit vectors pointing toward SNe $i$ and $j$, respectively, and ${\bf \betasn(r_i)}$, ${\bf \betasn(r_j)}$ are the corresponding SN velocity vectors. Since the velocity correlation function must be rotationally invariant, it can be decomposed in two components in terms of the comoving separation $r$ between SNe
\begin{equation}
\begin{aligned}
    \xi ({\bf r_i},{\bf{r_j}}) \,=\, &\sin \theta_i \sin \theta_j \xi_{\perp}(r, z_i,z_j) \\
    &+ \cos\theta_i 	\cos\theta_j \xi_{\parallel}(r,z_i,z_j) \,,
\end{aligned}
\end{equation}
with ${\bf r_{ij}} \equiv {\bf r_i} -{\bf r_j}$; $r = |{\bf r_{ij}} |$; $cos\theta_i \equiv \hat{{\bf r}_i} \cdot {\bf r_{ij}}$ and $cos \theta_j \equiv \hat{{\bf r}_j} \cdot {\bf r_{ij}}$. The parallel and perpendicular components of the velocity function are given by
\begin{equation}
    \xi_{\parallel,\perp} = G' (z_i) G' (z_j) \int^{ \infty}_0 \frac{\dd k}{2\pi^2}P(k)K_{\parallel,\perp}(kr)\,,
    \label{eq:xi}
\end{equation}
where $G$ is the growth function, $K_{\parallel}\equiv j_0(x)-2j_1(x)/x$, $K_{\perp}(x)\equiv j_1(x)/x$ and $j_0,j_1$ represent the spherical Bessel functions. For the diagonal $i=j$ terms, we have simply:
\begin{equation}
    \xi ({\bf r_i},{\bf{r_i}}) \,=\, \big[G'(z_i)\big]^2 \int^{ \infty}_0 \frac{\dd k}{2\pi^2}\frac{P(k)}{3}\,.
    \label{eq:xi-diag}
\end{equation}
The peculiar-motion covariance matrix is finally given by
\begin{equation}\label{eq:cov}
    C_{v}(i,j)=\left[1-\frac{(1+z_i)^2}{H(z_i)d_L(z_i)}\right] \bigg[(\dots)_{i \rightarrow j}\bigg] \xi ({\bf r_i},{\bf{r_j}}).
\end{equation}

Here, we extend the above usual analysis to include non-standard growth rates. This allows one to account for scenarios that go beyond the standard $\Lambda$CDM model. We employ the common parametrization for the linear growth rate $f$~\citep{Lahav:1991}
\begin{equation}\label{eq:gamma}
    f(z)\,=\,-\frac{\dd \ln G}{\dd \ln (1+z)} \simeq \Omega_m(z)^{\gamma}\,,
\end{equation}
where $\gamma$ is the growth rate index  and $\Omega_m(z) =\Omega_{m0} (1+z)^3 H_0^2/ H^2(z)$. Within General Relativity (GR) and for the standard $\Lambda$CDM model $f$ is accurately described by Eq.~\eqref{eq:gamma} with $\gamma=\gamma_{\rm \Lambda CDM} \approx 0.55$. It can be shown that the $\gamma$ parameter does not depend strongly on $w_{\rm DE}$, the equation of state parameter of dark energy~\citep{2010deto.book.....A} -- see also Fig.~7b of~\cite{Mantz:2014paa}. Therefore, $\gamma$ is often employed as a simple way of describing the growth-rate in modified gravity models.

\begin{figure}[t]
    \hspace{-.5cm}\includegraphics[width=0.53 \textwidth]
    {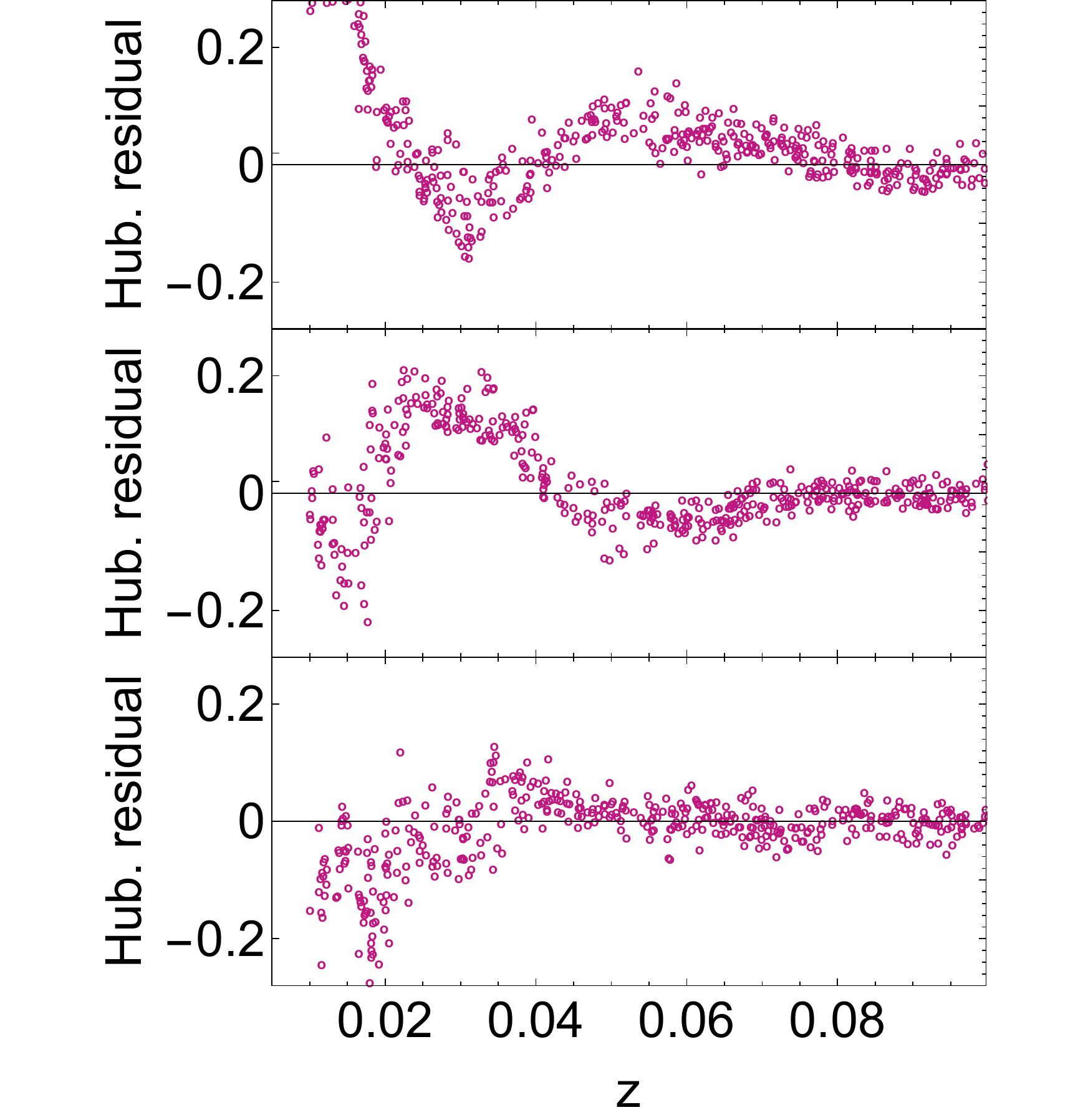}
    \caption{Illustration of the physical effect of velocities in the Hubble diagram residual for three random realizations. In all cases we assumed 500 idealized SNe (without intrinsic dispersion -- $\sigma_{\rm int} = 0$) in the range $0<z<0.1$ over a square 400 ${\rm deg}^2$ area. In these idealized cases the correlations are easily noticed by eye. In practice, these are suppressed by much larger intrinsic dispersion of the SNe.  \label{fig:PV-ideal}}
\end{figure}

Figure~\ref{fig:PV-ideal} illustrates the physical effect of the PV's in the Hubble diagram. Because of the PV, the Hubble residual (the difference between the SNe magnitudes and the best-fit cosmology) becomes correlated. In this figure, in order to make the effect clear by eye, we assume an idealized case where the SNe have no intrinsic dispersion, i.e. $\sigma_{\rm int} = 0$, and a scattered in a small area --- if the same SNe are distributed over a larger (smaller) area, the effect is diminished (enhanced). In particular, we show three random realizations generated using $500$ supernovae equally scattered both in redshift and in an area of the sky of 400 deg${}^2$. These are possible samples of these 500 SNe: we drew 3 samples of a 500 dimensional multi-normal distribution, with zero mean and a covariance matrix given by Eq.~\eqref{eq:cov} for a fiducial cosmological model with parameters close to the best-fit from Planck~\cite{Ade:2015xua}. To compute Eq.~\eqref{eq:cov} we employed the code made available in~\cite{hui2006}.\footnote{\url{http://www.astro.columbia.edu/~lhui/PairV/}}

\subsection{Likelihood parameters}\label{sec:parameters}
For the likelihood analysis, we used
\begin{equation}
    L_{\rm PV} \,\propto\, \frac{1}{\sqrt{|\Sigma_{\rm PV}|}} \,\exp\left[-\frac{1}{2} \delta_m^{T} \,\Sigma_{\rm PV} \,\delta_m \right]\,,
\end{equation}
where $\Sigma_{\rm PV}=C_{\rm JLA}(i,j) + C_{v}(i,j)$ is the full PV covariance matrix. The term $C_{\rm JLA}(i,j)$ is the covariance presented in Eq.~(13) of~\cite{Betoule:2014frx} (that combines terms from intrinsic dispersion of SN magnitudes, the uncertainty in the measurements from: statistics, calibration, model, bias, host, contamination, and dust; and degradation due to both lensing and peculiar velocities), minus two contributions: the peculiar velocity and intrinsic dispersion terms. The original JLA peculiar velocity covariance has two terms: one linear and one non-linear. The non-linear part is encapsulated in a simple parametrization: the diagonal matrix given by $(5\,\sigma_{\rm v})/(z\,\log\,10)$. The linear part is given by the matrix in~\eqref{eq:cov}. The intrinsic dispersion term originally had different values depending in which survey observed the supernova. Here, we replace it assuming a single value $\sigma_{\rm int}$ invoking Occam's Razor because we do not think these extra
parameters are needed. Furthermore, as can be seen in Fig.~(7) of~\cite{Betoule:2014frx}, a constant dispersion is a good fit in the range of our PV analysis.

The matter power spectrum was evaluated numerically using CAMB~\citep{Lewis:1999bs} in a 14-dimensional parameter space sampled by a Markov Chain Monte Carlo (MCMC) method. For cosmology we assume a model with the same background expansion as a flat $\Lambda$CDM and composed of 6 parameters: $\{\Omega_{c0}, \Omega_{b0}, \gamma, h, n_s, A\exp(-2\tau)\}$, where $h$ is $H_0$ in units of 100 km/(s.Mpc). These 6 are complemented by a set of 8 nuisance parameters: one is the SNe intrinsic magnitude $M$, three $(\alpha, \beta, \delta M)$ are from the standard SALT2 analysis, one $(\sigma_v)$ stands for a random component attributed to the non-linear velocity dispersions, two $(\sigma_{\rm int}^{\rm low\, z},\,\sigma_{\rm int}^{\rm hi\, z})$ for the intrinsic dispersion of the SNe, and one  $(\mu_{3,\,\rm int})$ for the intrinsic skewness of the SNe. These last three are important for the lensing analysis. From these 14 parameters we derive another two: $\Omega_{m0}\equiv\Omega_{c0}+ \Omega_{b0}$ and, from the
resulting matter power spectrum $P(k)$, the standard deviation of density perturbations on $8$ Mpc/h spheres:
\begin{equation}\label{eq:s8}
    \sigma_8\equiv \sqrt{\int{\dd k \frac{k^2}{2\pi^2} \frac{9\, P(k) }{(k R)^6}\big[ \sin(k R) - k R\cos(k R)  \big]^2}}\,,
\end{equation}
where $R\equiv 8$ Mpc/h.


It was shown in~\cite{Amendola:2013twa} that the lensing of SNe is not very sensitive to the value of a constant $w_{\rm DE}$. Likewise, as discussed above in Section~\ref{sec:pv-covar} the growth rate index $\gamma$ is also not very sensitive to $w_{\rm DE}$. We thus assume for simplicity $w_{\rm DE}\equiv w_{\Lambda}\equiv 1$ in our analysis.

Since SNe PV alone is not currently able to measure $h$, $n_s$ or $\Omega_{b0}$ with any significant sensitivity, we adopted the prior $h = 0.696 \pm 0.007$ from~\cite{Bennett:2014tka} and $\Omega_b h^2 = (0.021 \pm 0.01) $ from Big Bang Nucleosynthesis (BBN)~\cite{Iocco:2008va}). For $n_s$ we allowed a broad tophat in the range $[0.9-1.0]$. All other priors were assumed flat and much wider than the likelihood.  See~\ref{sec:MCMC} for more details on our MCMC implementation.

\section{Combined analysis of PV and lensing}\label{sec:combined}

Lensing on supernovae data was another effect that until recent works was merely treated as a nuisance systematic on supernovae magnitudes. The Method of the Moments (MeMo) was first presented in~\cite{Quartin:2013moa} and further enhanced in~\cite{Castro:2014oja}, where it was also tested with real data. It extends the traditional analysis to the higher moments of the Hubble diagram residual, and allows one to probe the matter properties intervening supernovae and observer. The MeMo analysis is basically a $\chi^2$ approach capable of quantifying the likelihood 
that a given sample was generated by a given observed PDF.

The important quantities of the MeMo are the sample moments of the distribution. It was shown by~\cite{Quartin:2013moa} that, concerning supernovae lensing, basically all the signal is contained in the first 4 moments (and most of it in the first 3). So, the MeMo likelihood of a supernovae sample distributed in a redshift bin $z_k$ is well approximated by:
\begin{equation}\label{eq:memo-lhood}
\begin{aligned}
    & L_{\rm{MeMo}}(z_k)\,=\,\frac{1}{(2\pi)^2 \sqrt{|\Sigma_{\rm MeMo}|}} \exp \left(-\frac{1}{2}\,\chi_{k}^{2}\right), \\
    & \chi_{k}^{2} \,=\, (\boldsymbol{\mu}-\boldsymbol{\mu}_{\rm{data}})^t \,\Sigma_{\rm MeMo}^{-1}\, (\boldsymbol{\mu}-\boldsymbol{\mu}_{\rm{data}}), \\
    & \boldsymbol{\mu}\,\equiv\,\{ {\mu_{1}^{\prime}},\mu_2,\mu_3,\mu_4 \}\,,
\end{aligned}
\end{equation}
where the $\{ \boldsymbol{\mu},\boldsymbol{\mu}_{\rm{data}}\}$ are the first moment and the second to fourth central moments of the observed PDF and the given sample respectively. The formula of the full covariance matrix $\Sigma_{\rm MeMo}$ involves all moments up to the $8^{\rm th}$, and the full equation was obtained in~\cite{Quartin:2013moa}. In order to compute it, we employ the \texttt{turboGL}\footnote{\href{http://www.turbogl.org/}{turbogl.org}} code, which implements the sGL method discussed in Section~\ref{sec:intro}.

The observed supernovae PDF is a result of the convolution of the lensing PDF with the supernovae intrinsic distribution. The lensing PDF was obtained using the sGL method discussed in Section~\ref{sec:intro}. Following~\cite{Castro:2014oja}, the supernovae intrinsic distribution was parametrized by two nuisance parameters: its intrinsic variance and skewness (respectively $\sigma_{\rm{int}}^2$ and $\mu_{\rm{3,int}}$). The degenerescence between the intrinsic moments and the lensing ones is broken due to the hypothesis that the intrinsic moments do not evolve with redshift. This hypothesis was strongly preferred by data in a Bayesian model selection and also agrees with the idea that the properties of a standard candle do not evolve with redshift. Therefore the observed SNe PDF moments are given by:
\begin{equation}
\begin{aligned}
    \mu_{\rm{2}}&\,=\,\mu_{\rm{2,lens}}+\sigma_{\rm{int}}^2 \,, \\
    \mu_{\rm{3}}&\,=\,\mu_{\rm{3,lens}}+\mu_{\rm{3,int}} \,, \\
    \mu_{\rm{4}}&\,=\,\mu_{\rm{4,lens}}+6\,\mu_{\rm{2,lens}}\,\sigma_{\rm{int}}^2
    +3\sigma_{\rm{int}}^4 \,.
\end{aligned}
\end{equation}

Regarding the sample moments, we suggested in~\cite{Castro:2014oja} that the best variable to compute the central moments is through the random variable:
\begin{equation}
    m_{j,k}\,=\,m_{j,k}^{\rm{catalog}}-m_{\rm{best}}(z_k)+m_{\rm{best}}(z_j),
\end{equation}
where $j$ and $k$ are the $j$-th supernova in the $k$-th bin. This is exactly as calculating the moments directly on the minimized Hubble residual diagram. Here we take a more robust approach and adapt this variable in order to include also the SALT-2 nuisance parameters. The supernovae magnitudes inferred through the SALT-2 light curve fitting are given by the following relation:
\begin{equation}
    m_{\rm{catalog}} = m_{B}^{*}-M-\Delta M +\alpha \, x -\beta \, c,
\end{equation}
where $m_{B}^{*}$, $x$ and $c$ are results from SALT-2 light curve fitting and $M$, $\Delta M$, $\alpha$ and $\beta$ are calculated minimizing the Hubble residual. Here, instead of using the moments calculated in the minimized Hubble residual, we calculate them as a function of the cosmological background parameters and the four SALT-2 nuisance parameters. Hence, our results of the MeMo analysis in this paper are slightly different from our previous work due to the marginalization over these nuisance parameters.

Note that although in principle $\Sigma_{\rm MeMo}$ depends on cosmology, \cite{Quartin:2013moa} showed that to good approximation one can fix the cosmology just keeping the redshift dependence. In that paper we did not have any light-curve fitting nuisance parameters, so here we compute it including them. So we have $\Sigma_{\rm MeMo}(z,\alpha,\beta,M,\Delta M)$.

The complementarity of the peculiar velocity (only sensitive to low-$z$ supernovae) and lensing (sensitive to medium to high-$z$ supernovae) signals motivates the analysis of these effects in a independent way. We therefore just need to compute two likelihoods, one for SNe in $z<0.1$ including the PV covariance, and one for SNe with $z\ge0.1$ including the higher moments, and multiply both to get the total constraints:
\begin{equation}
    L_{\rm{tot}}\,=\, L_{\rm{PV}} \, L_{\rm{MeMo}} \,.
\end{equation}

To illustrate the potential of such a combined analysis, we made a forecast for a possible future SNe catalog consisting of 3000 Dark Energy Survey (DES) supernovae, with the redshift distribution and error estimates made in~\cite{Bernstein:2011zf} (we adopt their ``hybrid5'' catalog), plus 1000 low-redshift SNe uniformly distributed in the range $0.01<z<0.1$ and in an area of the sky of 400 deg${}^2$, with an intrinsic scatter of $0.13$ mag. This is a very simplistic low-$z$ catalog. In reality, for a given telescope the distribution will be a combination of many factors: the fact that the observed volume grows as $z^2$, the cadence of observations, the integration time, etc. More realistic estimates will inevitably feature proportionally less events at very low $z$. But since the idea here with this forecast is more to illustrate the method, we leave a more detailed forecast for future investigations.

\begin{figure}[t]
    \includegraphics[width=0.47 \textwidth]{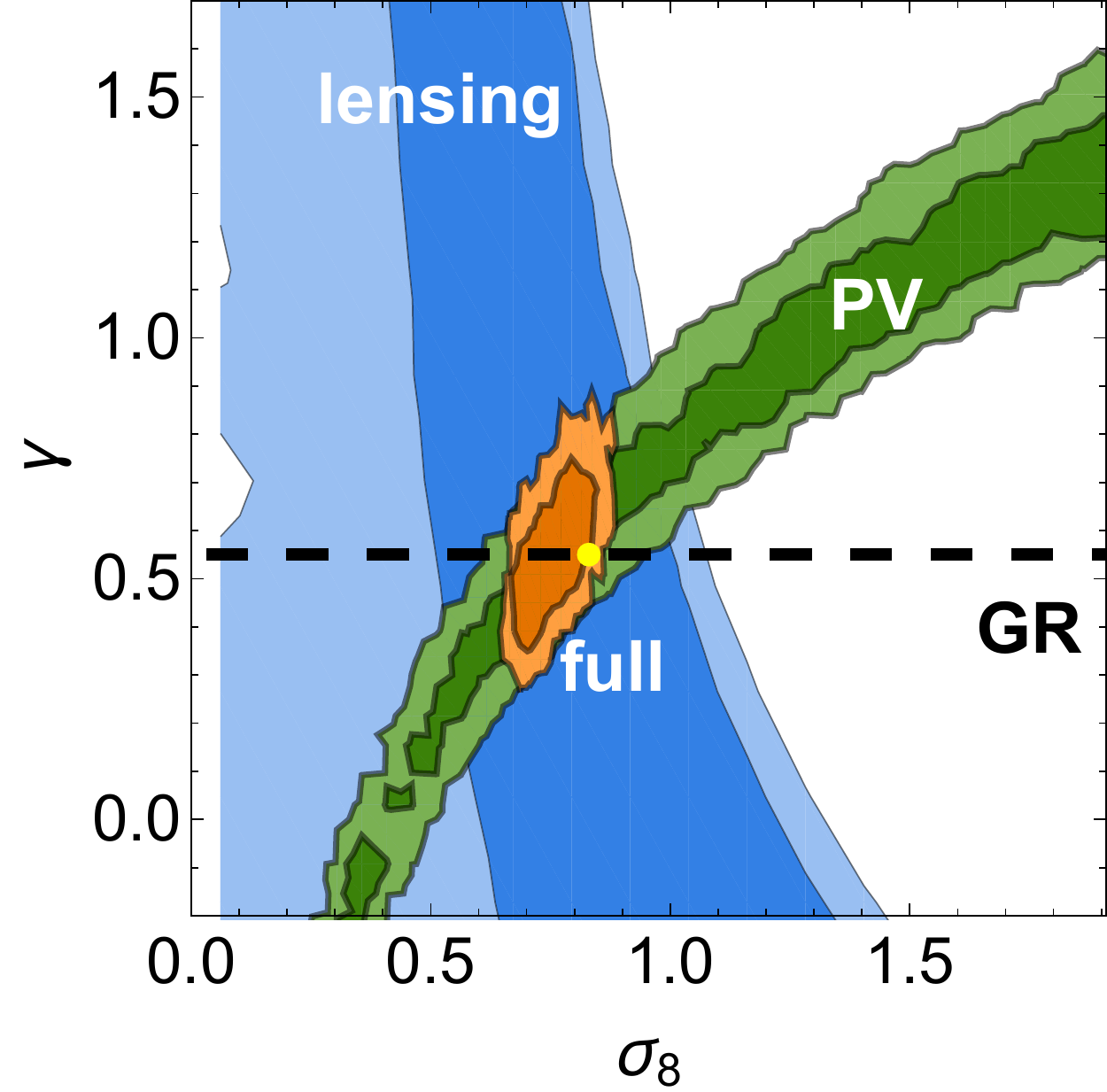}
    \caption{1 and $2\sigma$ forecast for a possible future SNe catalog consisting of 3000 DES plus 1000 low-redshift SNe uniformly distributed in the range $0.01<z<0.1$ and in an area of the sky of 400 deg${}^2$. We show the contours obtained by using only the peculiar velocities (PV) method  (green), using only the MeMo method (blue), or both combined (orange). The dashed black line represents the expectation from GR. Lensing and PV show good complementarity: the degeneracy lines make a $\sim 60^\circ$ angle around the fiducial point.  \label{fig:DES-mock}}
\end{figure}

Figure~\ref{fig:DES-mock} depicts the contours obtained with $L_{\rm{PV}}$, $L_{\rm{MeMo}}$ and $L_{\rm{tot}}$ separately. The combined contours provide a precise measurement of both  $\sigma_8$ and $\gamma$: the uncertainties are $\simeq 0.1$ and $\simeq 0.2$, respectively. Clearly, in this case most constraining power is coming from PV, so we plan to conduct in a future paper a careful analysis of the optimal observational strategy to detect SNe for measuring their PV correlations. As can be seen from the plot, the contours show good complementarity, and the degeneracy directions around the best fit subtend an angle of $\sim 60^\circ$. This nicely illustrates the benefit of combining different probes: the constraints coming solely from either PV or lensing exhibit pronounced parameter degeneration.

\section{Adjusting the JLA catalog for PV measurements}\label{sec:jla}

\begin{figure}[t]
    \centering
    \includegraphics[width=0.49 \textwidth]{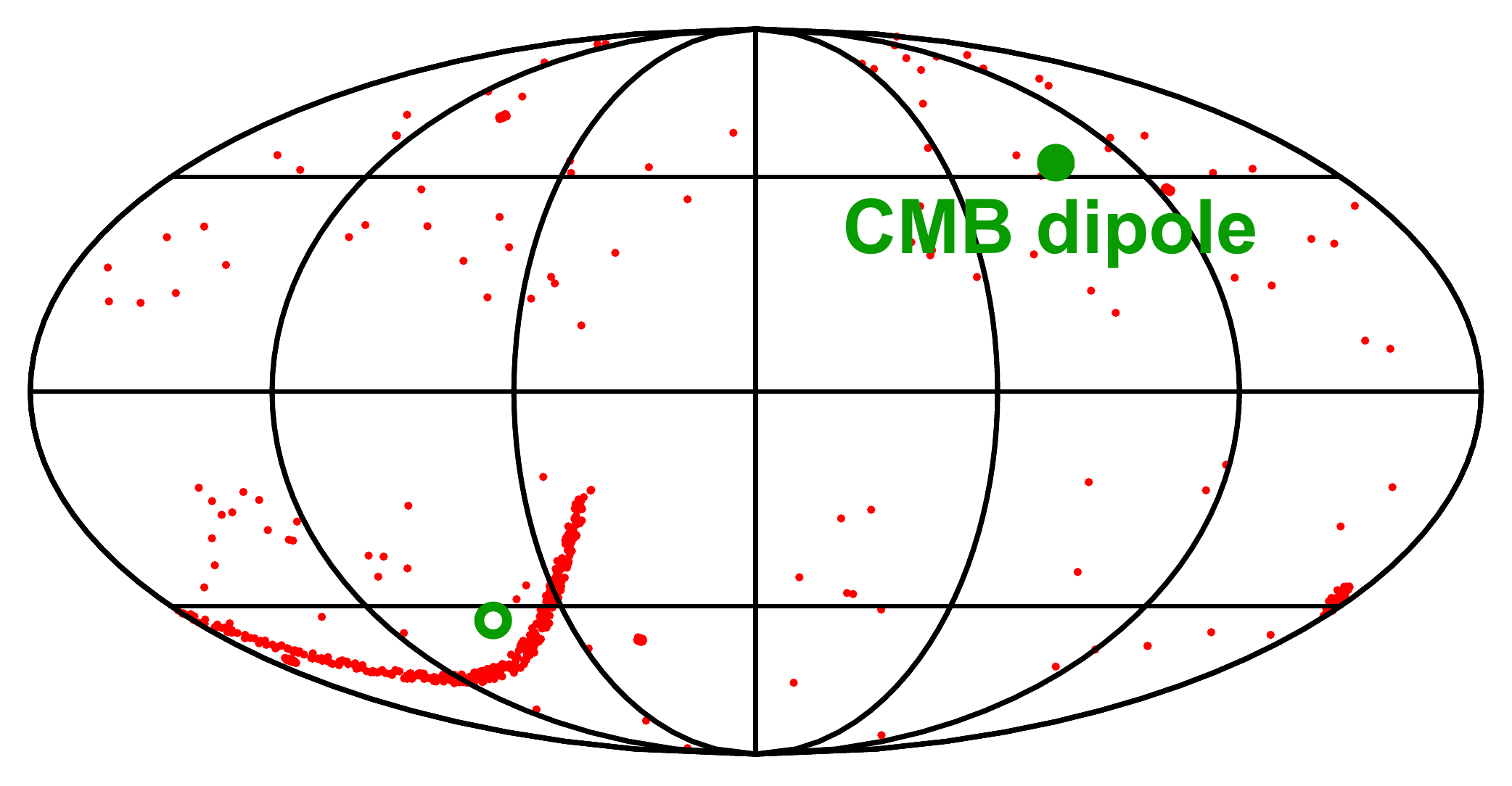}
    \includegraphics[width=0.47 \textwidth]{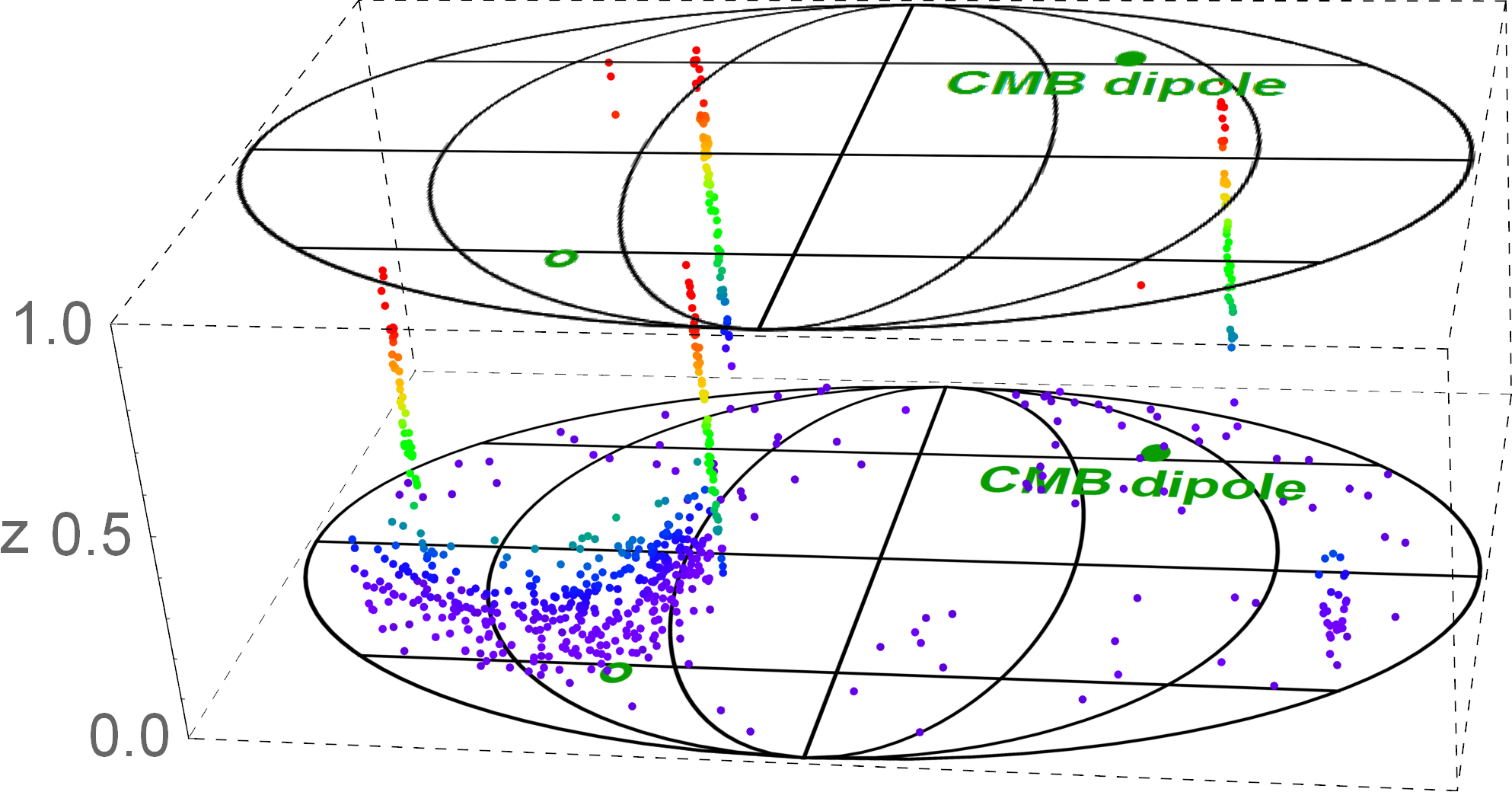}
    \caption{\emph{Top:} The spatial distribution of the JLA catalog in galactic coordinates, in the Mollweide projection. We also depict the CMB dipole direction with a full green circle and its antipode with an open circle. \emph{Bottom:} Similar plot but with a third dimension added to show also the redshift (with dots colored from purple to green to red as $z$ increases). Note that the catalog is mostly constituted of 4 deep fields, the mid-depth stripe-82 and low-$z$ SNe scattered around the sky. \label{fig:JLA}}
\end{figure}

In this paper we make use of the Joint-Lightcurve Analysis (JLA) sample~\citep{Betoule:2014frx} consisting of 740 objects. All SNe were spectroscopically confirmed and with high quality light-curve data, and light-curve calibration was performed using the SALT2 model of~\cite{Guy:2007dv} with significant improvements with respect to the previous analysis of~\cite{conley2011}. The dataset is well sampled in redshift up to $z\simeq 1$. The angular distribution of the SNe, on the other hand, is far from homogeneous. It is basically constituted of 4 deep fields, the mid-depth stripe-82 and a reasonably homogenous low-$z$ sample.  The spatial distribution of these 740 SNe are depicted in Figure~\ref{fig:JLA}. All our lensing signal therefore comes from these few deep fields, and most of our PV signal comes from the Stripe-82 due to the angular proximity of the SNe.

In the original JLA analysis the peculiar velocity model for the low-z sample was left unchanged with respect to the analysis of~\cite{conley2011}, where they correct for peculiar motion on a SN-by-SN basis. This approach follows the prescription of~\citet{hui2006} and consists on a two-step correction, first applied to redshifts through a slightly different version of Eq.~\eqref{eq:redshift_2}:
\begin{equation}
    (1+z_{\rm obs}) = (1+z_{\rm cosmo})(1+\betasn \cdot \nhat - \betao \cdot \nhat)\,,
    \label{eq:redshift_3}
\end{equation}
where $z_{\rm obs}$ is the observed redshift and $z_{\rm cosmo}$ is the redshift with respect to the CMB rest-frame.\footnote{We name it so because, aside any gravitational redshifts, it coincides with the pure cosmological expansion redshift.} The second step consists of a (much smaller) correction to the observed magnitude by adding the term $-5\log_{10}(1+\betasn \cdot \nhat)$.

Therefore in order to use the JLA catalog for peculiar velocities measurements it is better to undo the second step of this analysis to obtain the raw magnitudes (magnitudes that have the expected correlations between the supernovae). To undo this correction one needs $\betasn \cdot \nhat$, the component of the peculiar velocity of the supernovae in the line of sight of the observer. This can be done by direct inversion of Eq.~\eqref{eq:redshift_3}. JLA provides both $z_{\rm cosmo}$ and $z_{\rm obs}$, and $\betao \cdot \nhat$ is also known (assuming the kinematic explanation to the CMB dipole). We thus have:
\begin{equation}
    1+\betasn \cdot \nhat=\frac{1+z_{\rm obs}}{1+z_{\rm cosmo}}+\betao \cdot \nhat .
\end{equation}

A minor point has to be emphasized about how JLA catalog computes the $z_{\rm cosmo}$. To obtain this they correct the catalog accordingly to an peculiar velocity model, first presented in~\citet{Hudson:2004et}, and combine it with LSS data. For more details see~\citet{Betoule:2014frx} and references therein. Although this estimation uses other data and is based on a model, we believe it is warranted since it does not assume a cosmological model and shares the same basic assumptions as our analysis (basically the $\Lambda$CDM model). In any case, as will be pointed out in the end of this section, the correction on the magnitudes is sub-dominant.\footnote{The usage of $z_{\rm cosmo}$ in place of $z_{\rm cmb}$ (which is corrected only by our motion with respect to the CMB) also does not entice any statistically significant change.}

The raw magnitudes also have a dipole shift due to our own velocity in relation to the CMB rest frame. We have also removed this dipole using Eq.~\eqref{eq:dl_1}. Summarizing the magnitudes used in this analysis are given in relation of JLA magnitudes by:
\begin{equation}
\begin{aligned}
    m \,=\, & \;m_{\rm JLA}+5 \log_{10}(1+\betasn \cdot \nhat) \\
    &-5 \log_{10}(1-\betao \cdot \nhat)\,.
\end{aligned}
\end{equation}
This correction, although sub-dominant, is warranted as we understand that the peculiar velocity analysis has to be done in the observed magnitudes corrected only by our peculiar velocity. That being said, the crucial correction is on the redshifts of the supernovae, which are changed due our velocity in relation of CMB rest frame, $\betao$ -- see Eq.~\eqref{eq:redshift_2}.

Finally, as discussed in Section~\ref{sec:pecvel} one in principle should correct also the angular positions of the SNe due to the aberration effect. This affects the inferred separation of the SNe, but the overall effect is very small because if nearby objects share the same velocity flow, they will be equally aberrated and the effect on the correlation function will be zero. Moreover, most current SNe relevant for the PV signal happen to lie very close to the (antipode of) the CMB dipole direction (see Figure~\ref{fig:JLA}), where the aberration effect is at its minimum -- see Eq. (4) of~\cite{Amendola:2010ty}.

In any case, it is straightforward to remove the aberration induced by our own motion in the SNe catalog (the aberration induced by the SNe peculiar motion is trickier, because it requires an independent estimate of their velocity). Following~\cite{Challinor:2002zh}, the relation between the real position $\nhat^{\prime}$ of a given SNe and its observed position $\nhat$ is:
\begin{equation}
    \nhat^{\prime} = \frac{\nhat\cdot\vhat-\beta_{\rm o}}{1-\nhat\cdot\betao}\,\vhat + \frac{\left[\nhat-(\nhat\cdot\vhat)\vhat\right]}
    {(1- \nhat\cdot\betao)} \sqrt{1-\beta_{\rm o}^2} \,,
\end{equation}
where $\vhat \equiv \betao/ \beta_{\rm o}$ is the direction of our peculiar velocity (see Section~\ref{sec:intro}). We confirmed numerically that the aberration bias is indeed much smaller than current statistical uncertainties.

\section{Determination of $\sigma_8$ and $\gamma$}\label{sec:sigma}

As discussed in section~\ref{sec:combined}, lensing and PV are essentially uncorrelated. For the JLA catalog, in particular we have divided the sample into three uncorrelated redshift ranges: supernovae lying in the range $z \le 0.1$ were used in the peculiar velocity analysis;  supernovae within $0.1 < z \le 0.9$ were separated in 8 redshift bins with $\Delta z = 0.1$ and used for the MeMo analysis, and data above $z > 0.9$ were used only to evaluate the background cosmology,\footnote{The background analysis is also implicitly contained inside the PV and MeMo methods.} in the standard SALT-2 way. The reason for not using the MeMo for SNe with $z > 0.9$ is twofold. First, there are not many SNe in that range, thus compromising the MeMo method which relies on higher moments. In fact, for $z>1.0$ there are so few that the sample moments are mathematically ill-defined. For $0.9<z<1.0$ we could in principle apply the MeMo, but it was found in~\cite{Castro:2014oja} that this bin is an outlier in the method, and
including it decreases the goodness of fit. Since the origin of this discrepancy is still unknown, we also computed the likelihood including this bin in the MeMo analysis, but results did not change significantly, see~\ref{sec:tests}. The final MeMo likelihood is thus given by:
\begin{equation}
    L_{\rm{MeMo}}\,=\,\prod_{k=1}^{8}\frac{1}{(2\pi)^2 \sqrt{|\Sigma_{\rm MeMo}|}} \exp \left(-\frac{1}{2} \chi_{k}^{2} \right) ,
\end{equation}
and the total likelihood is
\begin{equation}
    L_{\rm{tot}}\,=\, L_{\rm{PV}} \, L_{\rm{MeMo}} \, L_{\rm{z\ge0.9}}\,.
\end{equation}

\begin{figure}[t]
    \includegraphics[width=0.47\textwidth]{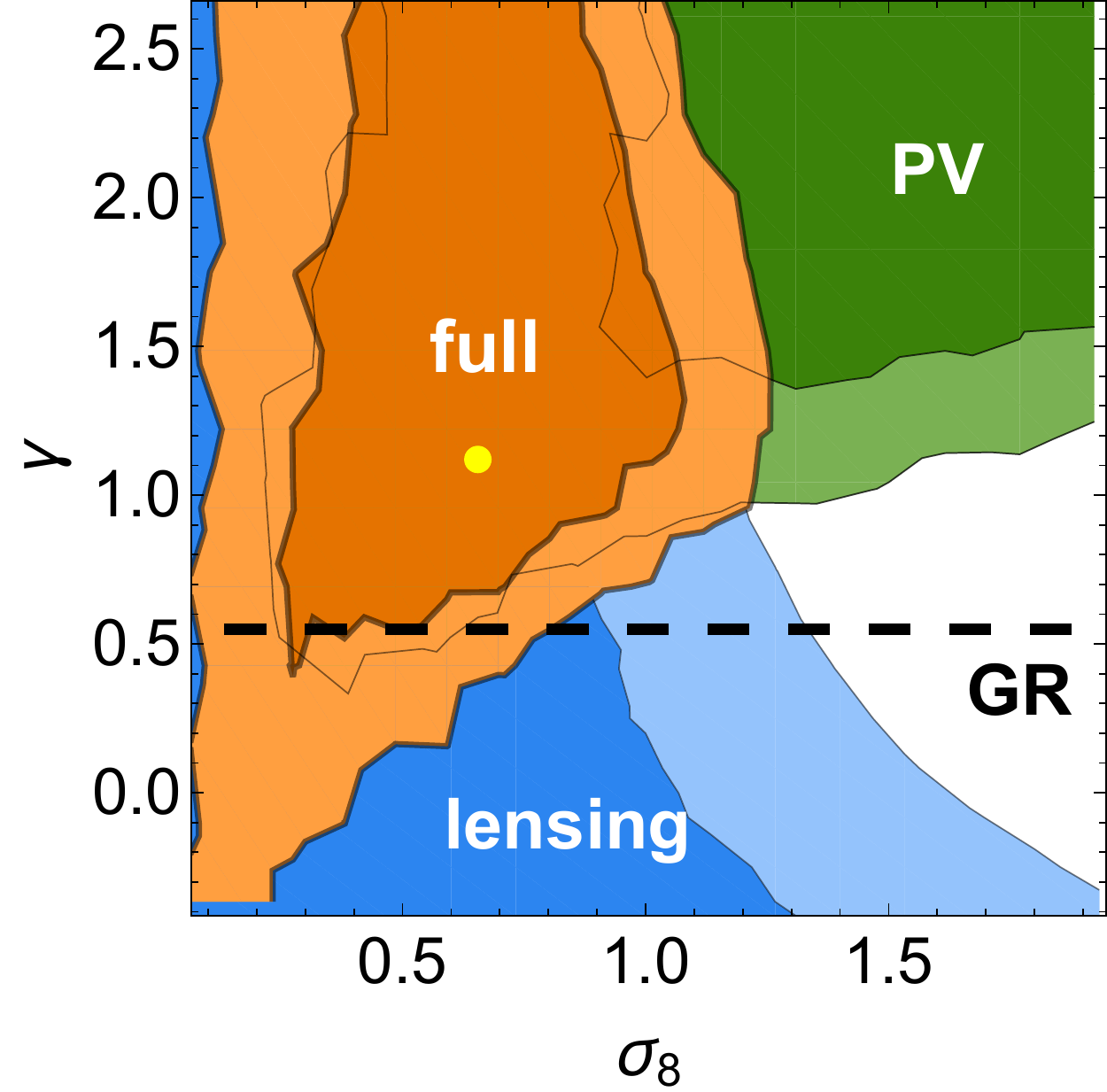}
    \caption{Same as Figure~\ref{fig:DES-mock} but for the real JLA data. Here, the yellow dot instead represents the best-fit point. Note that the PV contours are in $2\sigma$ tension with the expectations from CMB assuming GR, to wit $\sigma_8 \simeq 0.8$ and $\gamma=0.55$ \label{fig:PV+lens}}
\end{figure}

\begin{figure}[t]
    \includegraphics[width=0.47\textwidth]{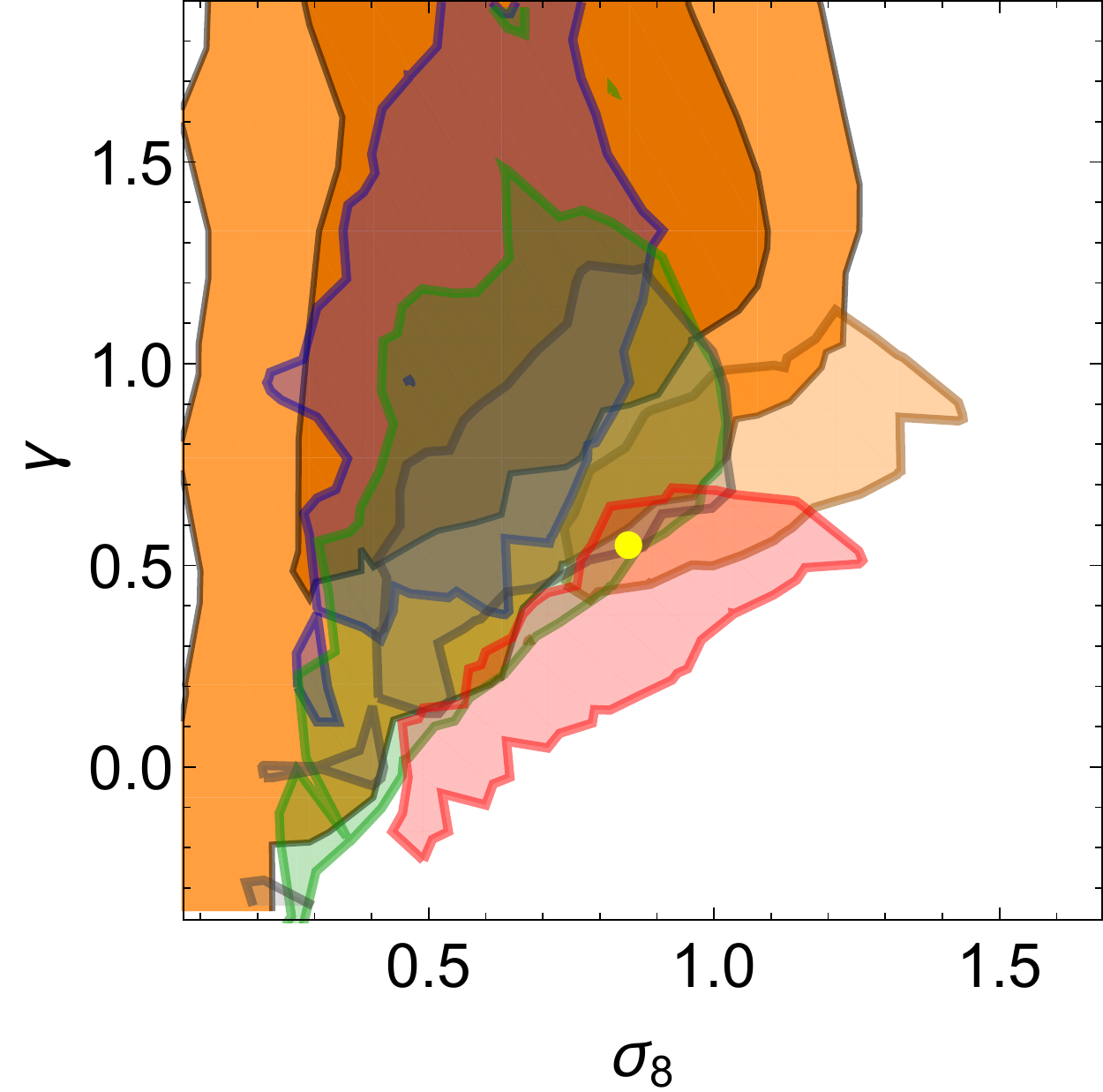}
    \caption{Simulations using mock catalogs of the combined constraints from SNe lensing and PV. In orange we repeat the the 1 and $2\sigma$ real data result of Figure~\ref{fig:PV+lens}. The 5 transparent $1\sigma$ contours are for 5 different random simulations. The yellow dot marks the fiducial value used in the simulations. Note that one of the 5 mocks produced contours very similar to the real data. \label{fig:mocks}}
\end{figure}

Figure \ref{fig:PV+lens} shows the two dimensional contours on $\{\sigma_8, \gamma\}$ for the JLA catalog. In blue are the contours concerning the MeMo analysis. In green we show the PV likelihood plus the background analysis. Finally, in orange is the combination of the two analysis (hereafter, dubbed simply ``full analysis''). As can be seen, the PV contours from the real data of the JLA catalog seem to be somewhat in tension with the expectations from CMB assuming GR, to wit $\sigma_8 \simeq 0.8$ and $\gamma=0.55$. In fact, along the $\gamma=0.55$ line, we see that PV favors a low $\sigma_8$, in agreement with the recent results of~\cite{Huterer:2015gpa}. Moreover, the PV contours are more vertically oriented than the forecasts with 4000 SNe, leading to only very loose constraints on $\gamma$. This may suggest that the JLA catalog has some unresolved systematic. We therefore conducted some simple tests, but could not find any hints pointing toward a culprit. We discuss these tests in~\ref{sec:tests}.

It is also important to note that the low statistical significance also means it could be just a random fluctuation. In order to not only test the amplitude of the statistical fluctuations on our combined analysis (and also our code), we conducted tests with mock catalogs. As the PV analysis is computationally intensive, we only investigated 5 such mocks. Each mock had the same spatial distribution of the JLA catalog and very similar statistical properties. More details of these catalogs are presented in~\ref{sec:mock-app}. Figure \ref{fig:mocks} shows the two dimensional contours on $\{\sigma_8, \gamma\}$ for these 5 mocks. They seem to be randomly distributed around the fiducial value $\{0.85, 0.55\}$ (presented as a yellow dot) without any clear bias.\footnote{We also conducted one mock test not including either PV or lensing, and we correctly inferred a value of $\sigma_8$ consistent with zero.} Interestingly, one of our 5 mocks (the one with blue contours) show a very similar shape to the real data,
again illustrating that the tension in the real data results could be a simple fluke. We also note that the other mocks were able to produce tighter constraints in $\gamma$. These large fluctuations in the constraining power should in any case be much smaller with more numerous supernovae catalogs.

\begin{figure}[t]
    \includegraphics[width=0.157\textwidth]{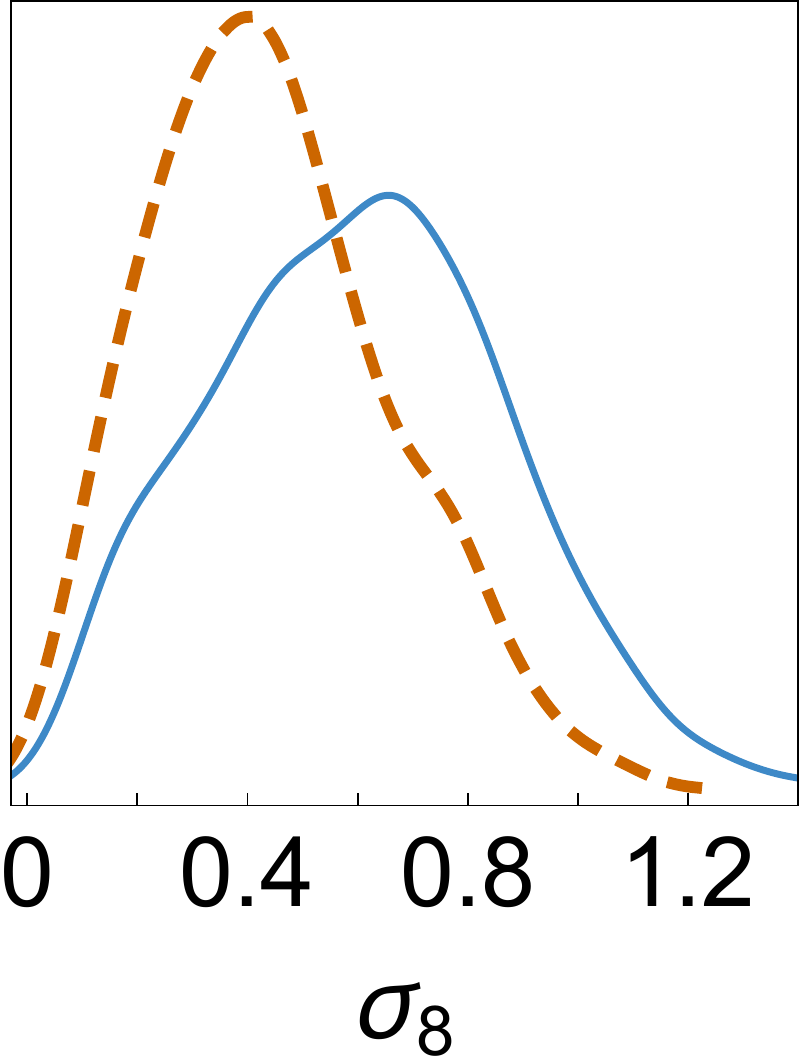}\!\!
    \includegraphics[width=0.158\textwidth]{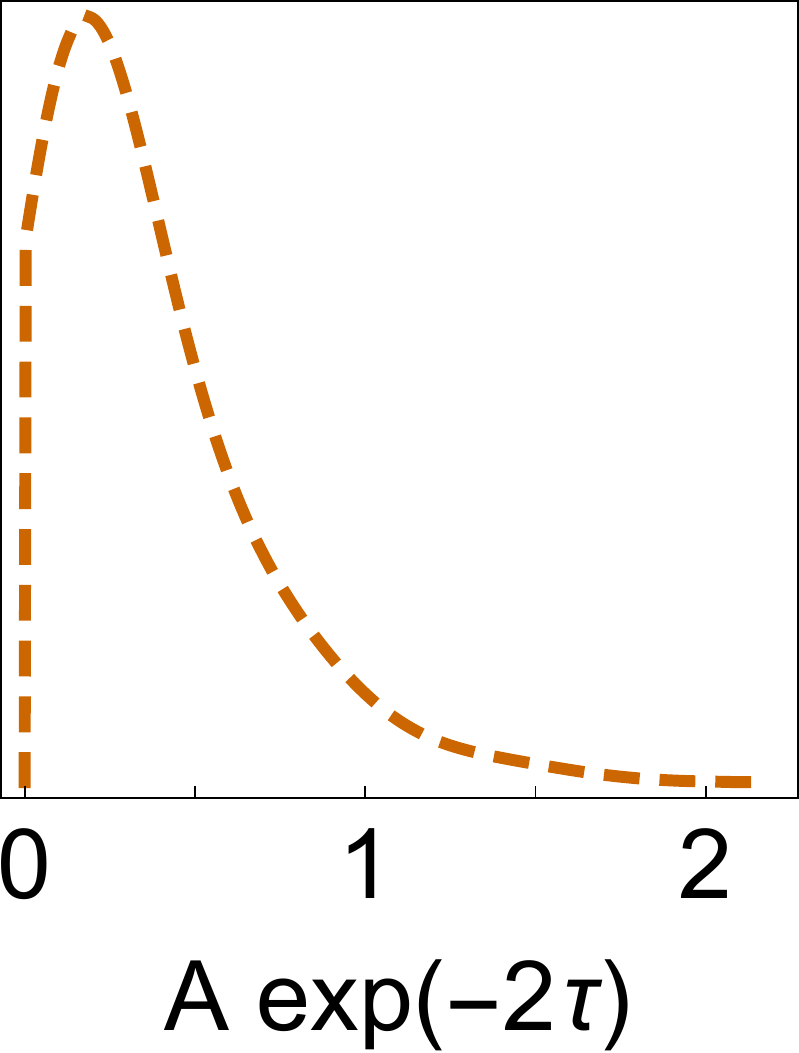}\!\!
    \includegraphics[width=0.158\textwidth]{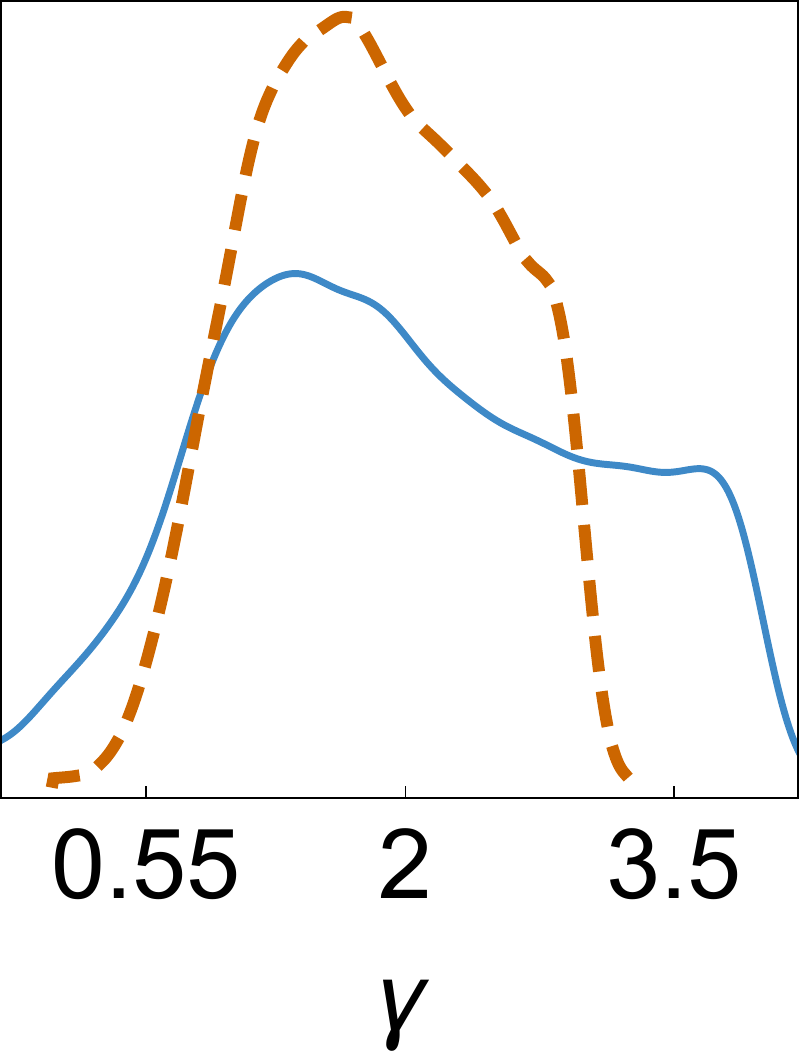}
    \caption{Marginalized 1-dimensional constraints on $\sigma_8$, $A \exp(-2\tau)$ and $\gamma$ from the combination of SNe lensing and peculiar velocities. In blue we show the full results and in dashed brown the contours obtained by fixing either $\gamma = 0.55$ (\emph{i.e.} assuming GR) [left and middle] or all perturbation quantities in $P(k)$ [right]. Note that for $\gamma \neq 0.55$ the variable $A \exp(-2\tau)$ is no longer directly related to the amplitude at the LSS, so we only plot the GR case contours for it.  \label{fig:1D-gamma-or-s8-fix}}
\end{figure}

Figure \ref{fig:1D-gamma-or-s8-fix} illustrates the results of our analysis if we fix the value of either $\gamma$ or the parameter set $\{\Omega_{b0}, \,H_0, \,A , \,n_s, \,\tau\}$, the ``perturbations'' parameters that enter the computation of $P(k)$. Although this is also true for $\Omega_{c0}$, we leave it free in this case otherwise we also fix the background expansion. Both analysis are well motivated: if we suppose that General Relativity is the correct theory of gravitation for cosmology, $\gamma$ is no longer a free parameter. On the other hand, if we rely on other precise measurements of the perturbation parameters, there is no difference in practice between marginalizing over the priors or fixing them to the best fit. We also show the posterior on the parameter $A \exp(-2\tau)$, from which we derive $\sigma_8$. The contours of $A \exp(-2\tau)$ are depicted in units of the Planck best fit, to wit $1.88 \times 10^{-9}$~\citep{Ade:2015xua}.
This can be directly compared with the recent results of~\citet{Huterer:2015gpa}: the results are similar even though we marginalize over many nuisance parameters.


\begin{figure*}[t]
    \centering
    \includegraphics[width=0.9\textwidth]{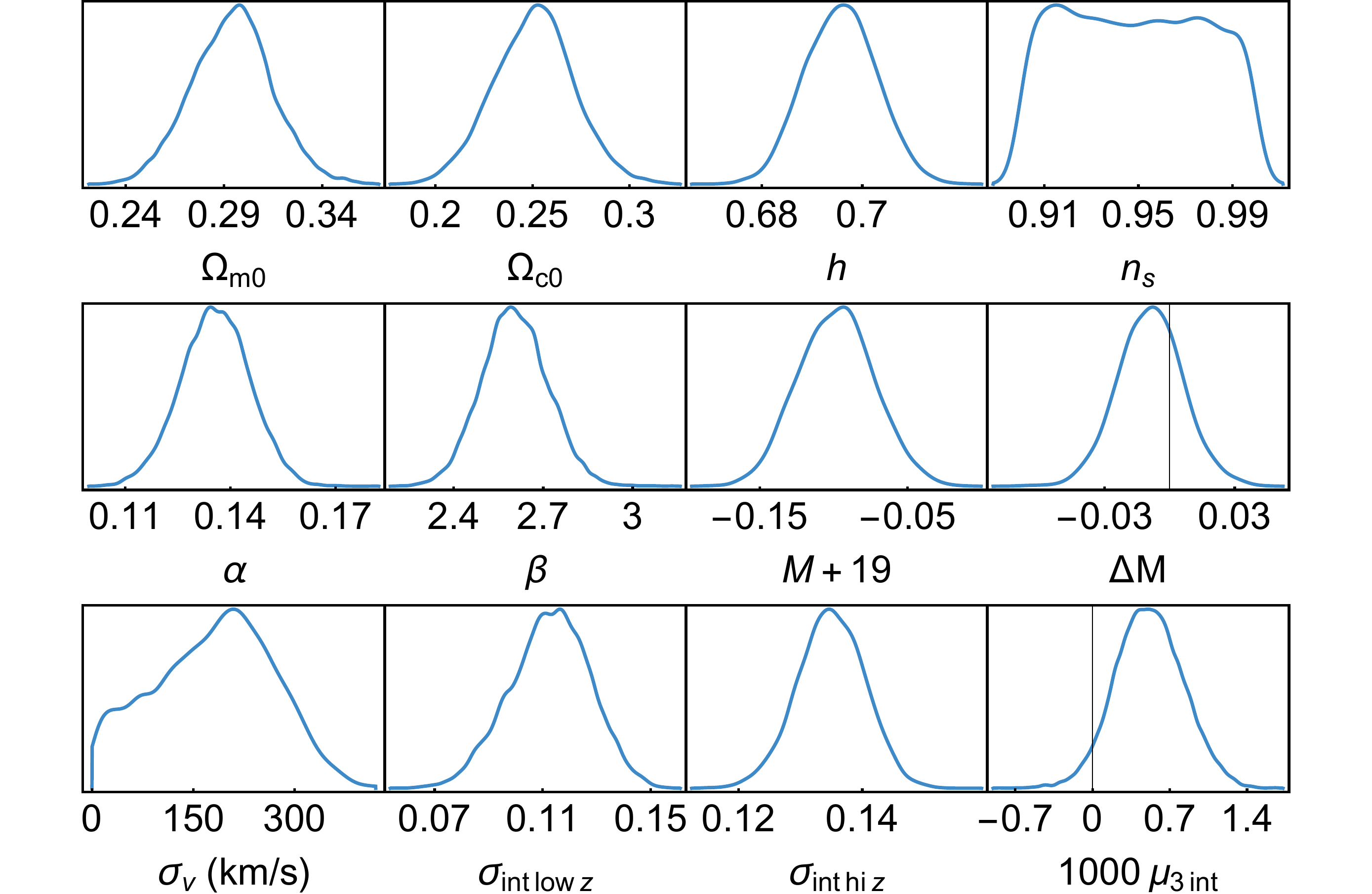}
    \caption{Marginalized 1-dimensional constraints from the combination of SNe lensing and peculiar velocities (PV). We only adopted tight priors for $h$ and $\Omega_{b0}$. For $n_s$ we used a top-hat in the region $[0.9, 1.0]$. All other variables had broad flat priors. The $\sigma_{\rm int}$ parameters are in mag, and $\mu_{3,{\rm int}}$ in mag${}^3$.} \label{fig:1D-lhoods}
\end{figure*}

In figure \ref{fig:1D-lhoods} we show all the one-dimensional constraints on the other twelve parameters. This figure exhibits the robustness of our results, illustrating that our chain covered all the high-likelihood region. More details of our MCMC analysis are provided in \ref{sec:MCMC}. In Table~\ref{tab:results} we summarize our main results: the measurements of $\sigma_8$ and $\gamma$. For the full analysis our best fit for $\sigma_8$ is in agreement with the values obtained with the CMB. However, traditional CMB analysis concerns the case of fixed $\gamma$, and in that case our best-fit is in roughly $2\sigma$ tension with the values obtained by Planck~\citep{Ade:2015xua}. We come back to this issue below when we compare the full contours of SNe and other datasets.

The values of $\gamma$ in Table~\ref{tab:results} for the full analysis are consistent with GR but the precision achieved is still worse compared to other results in literature: \cite{Mantz:2014paa}  found $\gamma = 0.48 \pm 0.19$ using only X-ray clusters number counts; \cite{Bocquet:2014lmj}  on the other hand obtained $\gamma = 0.73 \pm 0.28$ from a cluster sample selected in SPT survey  through the Sunyaev Zel'dovich effect. \cite{Abate:2008au} also used SNe PV to constrain $\gamma$ but did so fixing all other variables. They obtained $\gamma=0.72\pm0.21$, a result with much higher precision than ours in the case of fixed perturbation parameters, where although our constraints on $\gamma$ are tighter they are still over twice as large as their result (and in $2.3\sigma$ tension with GR). Finally, we note that JLA SNe have been used before to put constraints on $\gamma$ by~\cite{Amendola:2014yca} resulting in $\gamma = 0.52^{+0.16}_{-0.13}$, although in that paper this was driven mostly by the inclusion
of redshift-space distortions from a collection of galaxy surveys.

\begin{table}[t]
    \centering
    \begin{tabular}{ccccN}
        \hline
        & Full & Fixed pert.~vars. & $\,\;$Fixed $\gamma \;\,$ &\\
        \hline
        $\sigma_8$ & $0.65^{+0.23}_{-0.34}$ & $0.79\pm0.07$ (prior)& $0.40^{+0.21}_{-0.23}$ & \\[7pt]
        $\gamma $ & $1.38^{+1.7}_{-0.65}$  & $1.62^{+0.89}_{-0.55}$ & 0.55 (prior) &\\[7pt]
        \hline
    \end{tabular}
    \caption{Marginalized measurements of $\sigma_8$ and $\gamma$ from the combination of SNe lensing and peculiar velocities (PV). The first column is the general result; the second and third columns represent the cases in which we fixed the either $\gamma$ or all the perturbation variables in $P(k)$ to their fiducial values, together with the corresponding priors.  }
\label{tab:results}
\end{table}

\begin{figure}[t]
    \includegraphics[width=0.47\textwidth]{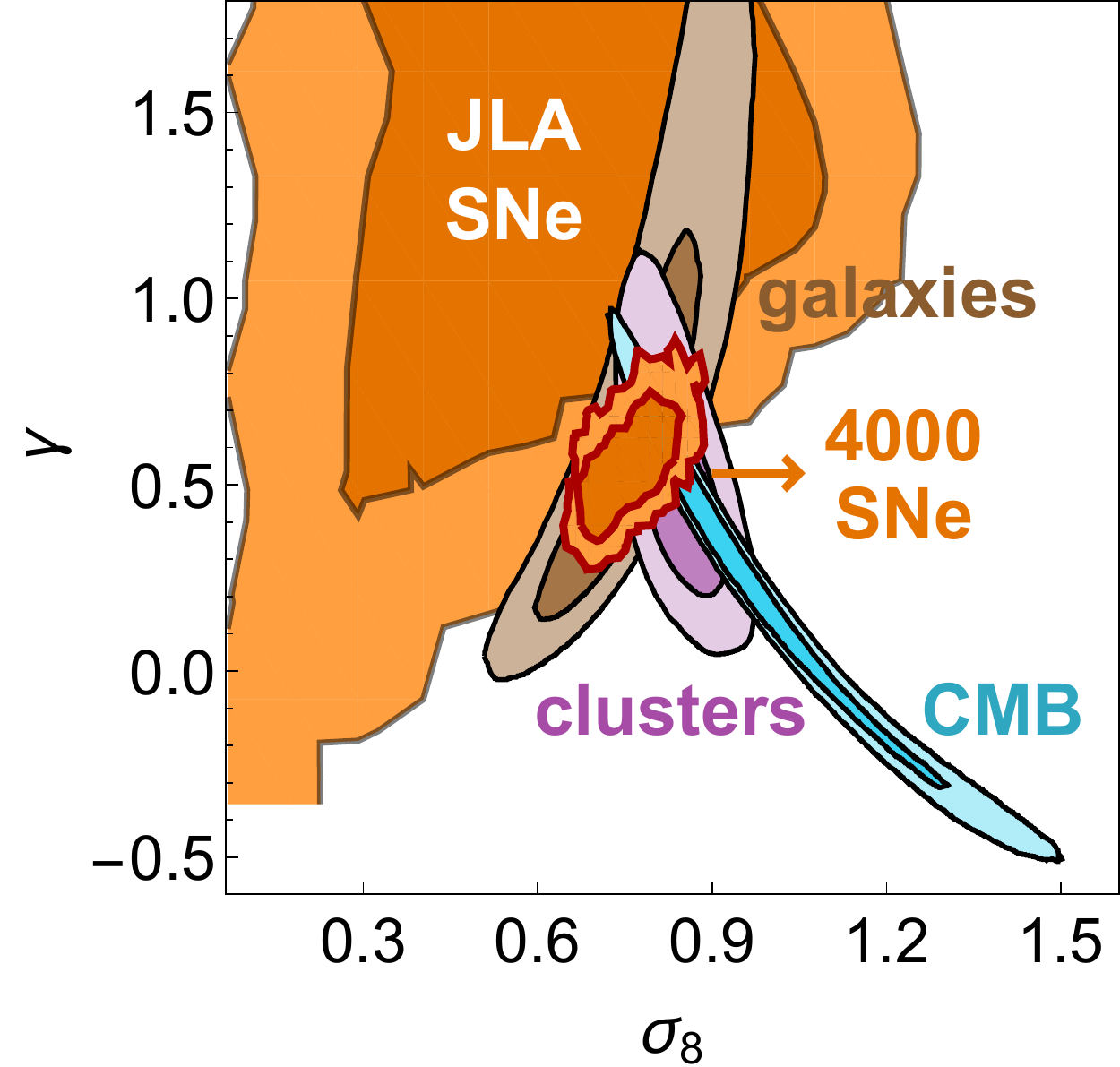}
    \caption{Comparison of the constraints on $\sigma_8$ and $\gamma$ from supernovae (using lensing and PV), Planck CMB spectra, clusters and galaxies. The large orange contours are for the JLA supernovae, whereas the smaller orange contours with red borders are a forecast for DES + 1000 low-z SNe, as discussed in Section~\ref{sec:combined}.
    The non-SNe contours were obtained by~\cite{Mantz:2014paa}. Note that current SNe are already able to complement well the other data, specially the CMB. \label{fig:SNe-vs-other-data}}
\end{figure}

We depict in Figure~\ref{fig:SNe-vs-other-data} a direct comparison of the SNe constraints with constraints from 3 other datasets, as obtained in~\cite{Mantz:2014paa}. The three external datasets consist of Planck CMB power spectra (complemented by large-scale WMAP polarization), X-ray selected cluster masses and galaxy surveys (both redshift-space distortions and Alcock--Paczynski effects). We refer the reader to~\cite{Mantz:2014paa} for more details. In order to construct this plot, we reverse engineered the contours from Fig. B1 of their paper, interpolating the overlapped regions. Although current SNe data do not provide competitive constraints, it is interesting to see that even in the near future\footnote{DES is already commissioning and a couple of low-z SN project releases are expected for the next years, such as Supernovae Factory~\citep{Aldering:2006uy}, Carnegie Supernovae Project~\citep{Hamuy:2005tf}, La Silla/QUEST~\citep{2012IAUS..285..324H}, SkyMapper~\citep{Keller:2007xt}, and Palomar
Transient Factory~\citep{Law:2009ys}.} the statistical power on SNe  will be very competitive and complement well for instance the Planck CMB contours, which in fact only achieve the quoted percent-level measurements of~$\sigma_8$~\citep{Ade:2015xua} by assuming $\gamma \equiv 0.55$.

We find the results here presented to be specially interesting because they show that SNe data alone will be capable in the near-future of constraining $\sigma_8$ and $\gamma$, in addition to the standard background quantities. Moreover, the results obtained with current data may indicate that a systematic is still unresolved, and our analysis is exercising a nice cross-check role. It is worth noting that we carried here a conservative analysis, in the sense that our results were marginalized over all nuisance parameters which are often left fixed in literature.

\section{Conclusions}\label{sec:conc}

In this paper we present current constraints and future forecasts on the matter power spectrum and growth of structure using supernovae data only. To do so, we split the JLA supernovae catalog in three redshift ranges. Nearby supernovae ($z<0.1$) were analyzed evaluating their peculiar velocity correlations (described in Section~\ref{sec:pecvel}). Supernovae farther away ($0.1<z<0.9$) were analyzed extracting their weak-lensing induced non-Gaussianity distribution (see Section~\ref{sec:intro}). Finally, for the few supernovae with $z>0.9$ (which apparently are not well modeled by our lensing method), the standard background analysis was adopted. In our analysis we marginalized over all parameters of ignorance, including the light-curve fitting ones. We thus take them to be very robust.

The peculiar velocity and lensing analysis are both sensitive to the power spectrum and  its evolution, but on their own both have pronounced degeneracies between not only these parameters but also with the background ones. On the other hand, the standard SN analysis are completely insensitive to any perturbation parameters and just constrain the background cosmology. We showed that in the near-future by combining these three probes not only all these degeneracies will be broken, as can be seen in Figure~\ref{fig:DES-mock}, but also competitive constraints will be achieved due to an almost orthogonal degeneracy between lensing and PV contours.

Our results for the JLA catalog were not as expected. The PV analysis produced results in roughly $2\sigma$ tension with expectations from CMB assuming GR. On the other hand, similar results were also obtained with 1 of our 5 mock catalogs. A robust assessment of this tension would require running at the very least dozens of such mock analysis, but this is a computationally intensive task and we did not have the resources to execute. In any case, we also see that some of the other JLA mocks produced much tighter constraints in $\gamma$, so if the current tension is not due to unknown systematics, future SN catalogs have the ability to put much tighter constraints in $\gamma$.

In closing, the combination of PV and lensing opens a new possibility of analysis capable of constraining both the present matter power spectrum and its growth using only supernova data. This combined method has a great potential in the future as the largest current catalog still constitutes but a small fraction of the total number of SN explosions that occur every year.

\section*{Note Added}

In the analysis in the original pre-print version of this manuscript we had a bug in our code [a simple sign error in $\betao$ in eq.~\eqref{eq:redshift_2}] that coincidentally shifted our results to be more in agreement with expectations from GR and CMB (and also with one of our mock catalogs). Since $\betao$ was not included in our mocks, their results were not affected by this bug and we thus missed it.

\section*{Acknowledgments}

We would like to thank Alex Conley for providing helpful insights and clarifications on the JLA data and Dragan Huterer, Daniel Shafera and Fabian Schmidt for pointing out imprecisions in the original version of this manuscript: this motivated us to search and find a bug in our code. It is also a pleasure to acknowledge Beatriz Siffert, Maurício Calvão and Ribamar Reis for fruitful discussions. TC is supported by the Brazilian research agency CAPES. MQ is supported by Brazilian research agencies CNPq and FAPERJ. SB is supported by Brazilian research agency CNPq.

\vspace{4mm}

\bibliographystyle{elsarticle-harv}
\bibliography{references}

\begin{thebibliography}{64}
\expandafter\ifx\csname natexlab\endcsname\relax\def\natexlab#1{#1}\fi
\providecommand{\url}[1]{\texttt{#1}}
\providecommand{\href}[2]{#2}
\providecommand{\path}[1]{#1}
\providecommand{\DOIprefix}{doi:}
\providecommand{\ArXivprefix}{arXiv:}
\providecommand{\URLprefix}{URL: }
\providecommand{\Pubmedprefix}{pmid:}
\providecommand{\doi}[1]{\href{http://dx.doi.org/#1}{\path{#1}}}
\providecommand{\Pubmed}[1]{\href{pmid:#1}{\path{#1}}}
\providecommand{\bibinfo}[2]{#2}
\ifx\xfnm\relax \def\xfnm[#1]{\unskip,\space#1}\fi
\bibitem[{Abate and Lahav(2008)}]{Abate:2008au}
\bibinfo{author}{Abate, A.}, \bibinfo{author}{Lahav, O.}, \bibinfo{year}{2008}.
\newblock \bibinfo{title}{{The Three Faces of $\Omega_m$: Testing Gravity with
  Low and High Redshift SN Ia Surveys}}.
\newblock \bibinfo{journal}{Mon. Not. Roy. Astron. Soc.} \bibinfo{volume}{389},
  \bibinfo{pages}{47}.
\newblock \DOIprefix\doi{10.1111/j.1745-3933.2008.00519.x},
  \href{http://arxiv.org/abs/0805.3160}{{\tt arXiv:0805.3160}}.
\bibitem[{Abell et~al.(2009)}]{Abell:2009aa}
\bibinfo{author}{Abell, P.A.}, et~al. (\bibinfo{collaboration}{LSST Science
  Collaborations, LSST Project}), \bibinfo{year}{2009}.
\newblock \bibinfo{title}{{LSST Science Book, Version 2.0}}
  \href{http://arxiv.org/abs/0912.0201}{{\tt arXiv:0912.0201}}.
\bibitem[{Adam et~al.(2015)}]{Adam:2015vua}
\bibinfo{author}{Adam, R.}, et~al. (\bibinfo{collaboration}{Planck}),
  \bibinfo{year}{2015}.
\newblock \bibinfo{title}{{Planck 2015 results. VIII. High Frequency Instrument
  data processing: Calibration and maps}}
  \href{http://arxiv.org/abs/1502.01587}{{\tt arXiv:1502.01587}}.
\bibitem[{Ade et~al.(2015)}]{Ade:2015xua}
\bibinfo{author}{Ade, P.A.R.}, et~al. (\bibinfo{collaboration}{Planck}),
  \bibinfo{year}{2015}.
\newblock \bibinfo{title}{{Planck 2015 results. XIII. Cosmological parameters}}
  \href{http://arxiv.org/abs/1502.01589}{{\tt arXiv:1502.01589}}.
\bibitem[{Aghanim et~al.(2014)}]{Aghanim:2013suk}
\bibinfo{author}{Aghanim, N.}, et~al. (\bibinfo{collaboration}{Planck}),
  \bibinfo{year}{2014}.
\newblock \bibinfo{title}{{Planck 2013 results. XXVII. Doppler boosting of the
  CMB: Eppur si muove}}.
\newblock \bibinfo{journal}{Astron.Astrophys.} \bibinfo{volume}{571},
  \bibinfo{pages}{A27}.
\newblock \DOIprefix\doi{10.1051/0004-6361/201321556},
  \href{http://arxiv.org/abs/1303.5087}{{\tt arXiv:1303.5087}}.
\bibitem[{Aldering et~al.(2006)}]{Aldering:2006uy}
\bibinfo{author}{Aldering, G.}, et~al. (\bibinfo{collaboration}{Nearby
  Supernova Factory}), \bibinfo{year}{2006}.
\newblock \bibinfo{title}{{Nearby Supernova Factory Observations of SN 2005gj:
  Another Type Ia Supernova in a Massive Circumstellar Envelope}}.
\newblock \bibinfo{journal}{Astrophys. J.} \bibinfo{volume}{650},
  \bibinfo{pages}{510--527}.
\newblock \DOIprefix\doi{10.1086/507020},
  \href{http://arxiv.org/abs/astro-ph/0606499}{{\tt arXiv:astro-ph/0606499}}.
\bibitem[{Amendola et~al.(2015)Amendola, Castro, Marra and
  Quartin}]{Amendola:2014yca}
\bibinfo{author}{Amendola, L.}, \bibinfo{author}{Castro, T.},
  \bibinfo{author}{Marra, V.}, \bibinfo{author}{Quartin, M.},
  \bibinfo{year}{2015}.
\newblock \bibinfo{title}{{Constraining the growth of perturbations with
  lensing of supernovae}}.
\newblock \bibinfo{journal}{Mon. Not. Roy. Astron. Soc.} \bibinfo{volume}{449},
  \bibinfo{pages}{2845--2852}.
\newblock \DOIprefix\doi{10.1093/mnras/stv497},
  \href{http://arxiv.org/abs/1412.3703}{{\tt arXiv:1412.3703}}.
\bibitem[{Amendola et~al.(2011)Amendola, Catena, Masina, Notari, Quartin and
  Quercellini}]{Amendola:2010ty}
\bibinfo{author}{Amendola, L.}, \bibinfo{author}{Catena, R.},
  \bibinfo{author}{Masina, I.}, \bibinfo{author}{Notari, A.},
  \bibinfo{author}{Quartin, M.}, \bibinfo{author}{Quercellini, C.},
  \bibinfo{year}{2011}.
\newblock \bibinfo{title}{{Measuring our peculiar velocity on the CMB with
  high-multipole off-diagonal correlations}}.
\newblock \bibinfo{journal}{JCAP} \bibinfo{volume}{1107}, \bibinfo{pages}{027}.
\newblock \DOIprefix\doi{10.1088/1475-7516/2011/07/027},
  \href{http://arxiv.org/abs/1008.1183}{{\tt arXiv:1008.1183}}.
\bibitem[{Amendola et~al.(2010)Amendola, Kainulainen, Marra and
  Quartin}]{Amendola:2010ub}
\bibinfo{author}{Amendola, L.}, \bibinfo{author}{Kainulainen, K.},
  \bibinfo{author}{Marra, V.}, \bibinfo{author}{Quartin, M.},
  \bibinfo{year}{2010}.
\newblock \bibinfo{title}{{Large-scale inhomogeneities may improve the cosmic
  concordance of supernovae}}.
\newblock \bibinfo{journal}{Phys.Rev.Lett.} \bibinfo{volume}{105},
  \bibinfo{pages}{121302}.
\newblock \DOIprefix\doi{10.1103/PhysRevLett.105.121302},
  \href{http://arxiv.org/abs/1002.1232}{{\tt arXiv:1002.1232}}.
\bibitem[{{Amendola} and {Tsujikawa}(2010)}]{2010deto.book.....A}
\bibinfo{author}{{Amendola}, L.}, \bibinfo{author}{{Tsujikawa}, S.},
  \bibinfo{year}{2010}.
\newblock \bibinfo{title}{{Dark Energy: Theory and Observations}}.
\newblock \bibinfo{publisher}{Cambridge University Press}.
\bibitem[{Bahcall et~al.(1999)Bahcall, Ostriker, Perlmutter and
  Steinhardt}]{Bahcall:1999xn}
\bibinfo{author}{Bahcall, N.A.}, \bibinfo{author}{Ostriker, J.P.},
  \bibinfo{author}{Perlmutter, S.}, \bibinfo{author}{Steinhardt, P.J.},
  \bibinfo{year}{1999}.
\newblock \bibinfo{title}{{The Cosmic triangle: Assessing the state of the
  universe}}.
\newblock \bibinfo{journal}{Science} \bibinfo{volume}{284},
  \bibinfo{pages}{1481--1488}.
\newblock \DOIprefix\doi{10.1126/science.284.5419.1481},
  \href{http://arxiv.org/abs/astro-ph/9906463}{{\tt arXiv:astro-ph/9906463}}.
\bibitem[{Ben-Dayan et~al.(2013)Ben-Dayan, Gasperini, Marozzi, Nugier and
  Veneziano}]{BenDayan:2013gc}
\bibinfo{author}{Ben-Dayan, I.}, \bibinfo{author}{Gasperini, M.},
  \bibinfo{author}{Marozzi, G.}, \bibinfo{author}{Nugier, F.},
  \bibinfo{author}{Veneziano, G.}, \bibinfo{year}{2013}.
\newblock \bibinfo{title}{{Average and dispersion of the luminosity-redshift
  relation in the concordance model}}.
\newblock \bibinfo{journal}{JCAP} \bibinfo{volume}{1306}, \bibinfo{pages}{002}.
\newblock \DOIprefix\doi{10.1088/1475-7516/2013/06/002},
  \href{http://arxiv.org/abs/1302.0740}{{\tt arXiv:1302.0740}}.
\bibitem[{Bennett et~al.(2014)Bennett, Larson, Weiland and
  Hinshaw}]{Bennett:2014tka}
\bibinfo{author}{Bennett, C.L.}, \bibinfo{author}{Larson, D.},
  \bibinfo{author}{Weiland, J.L.}, \bibinfo{author}{Hinshaw, G.},
  \bibinfo{year}{2014}.
\newblock \bibinfo{title}{{The 1\% Concordance Hubble Constant}}.
\newblock \bibinfo{journal}{Astrophys. J.} \bibinfo{volume}{794},
  \bibinfo{pages}{135}.
\newblock \DOIprefix\doi{10.1088/0004-637X/794/2/135},
  \href{http://arxiv.org/abs/1406.1718}{{\tt arXiv:1406.1718}}.
\bibitem[{Bernardeau et~al.(1997)Bernardeau, Van~Waerbeke and
  Mellier}]{Bernardeau:1996un}
\bibinfo{author}{Bernardeau, F.}, \bibinfo{author}{Van~Waerbeke, L.},
  \bibinfo{author}{Mellier, Y.}, \bibinfo{year}{1997}.
\newblock \bibinfo{title}{{Weak lensing statistics as a probe of Omega and
  power spectrum}}.
\newblock \bibinfo{journal}{Astron.Astrophys.} \bibinfo{volume}{322},
  \bibinfo{pages}{1--18}.
\newblock \href{http://arxiv.org/abs/astro-ph/9609122}{{\tt
  arXiv:astro-ph/9609122}}.
\bibitem[{Bernstein et~al.(2012)Bernstein, Kessler, Kuhlmann, Biswas, Kovacs
  et~al.}]{Bernstein:2011zf}
\bibinfo{author}{Bernstein, J.}, \bibinfo{author}{Kessler, R.},
  \bibinfo{author}{Kuhlmann, S.}, \bibinfo{author}{Biswas, R.},
  \bibinfo{author}{Kovacs, E.}, et~al., \bibinfo{year}{2012}.
\newblock \bibinfo{title}{{Supernova Simulations and Strategies For the Dark
  Energy Survey}}.
\newblock \bibinfo{journal}{Astrophys.J.} \bibinfo{volume}{753},
  \bibinfo{pages}{152}.
\newblock \DOIprefix\doi{10.1088/0004-637X/753/2/152},
  \href{http://arxiv.org/abs/1111.1969}{{\tt arXiv:1111.1969}}.
\bibitem[{Betoule et~al.(2014)}]{Betoule:2014frx}
\bibinfo{author}{Betoule, M.}, et~al. (\bibinfo{collaboration}{SDSS
  Collaboration}), \bibinfo{year}{2014}.
\newblock \bibinfo{title}{{Improved cosmological constraints from a joint
  analysis of the SDSS-II and SNLS supernova samples}}.
\newblock \bibinfo{journal}{Astron.Astrophys.} \bibinfo{volume}{568},
  \bibinfo{pages}{A22}.
\newblock \DOIprefix\doi{10.1051/0004-6361/201423413},
  \href{http://arxiv.org/abs/1401.4064}{{\tt arXiv:1401.4064}}.
\bibitem[{{Blake} et~al.(2011)}]{blake2011}
\bibinfo{author}{{Blake}, C.}, et~al., \bibinfo{year}{2011}.
\newblock \bibinfo{title}{{The WiggleZ Dark Energy Survey: mapping the
  distance-redshift relation with baryon acoustic oscillations}}.
\newblock \bibinfo{journal}{\mnras} \bibinfo{volume}{418},
  \bibinfo{pages}{1707--1724}.
\newblock \DOIprefix\doi{10.1111/j.1365-2966.2011.19592.x},
  \href{http://arxiv.org/abs/1108.2635}{{\tt arXiv:1108.2635}}.
\bibitem[{Bocquet et~al.(2015)}]{Bocquet:2014lmj}
\bibinfo{author}{Bocquet, S.}, et~al. (\bibinfo{collaboration}{SPT}),
  \bibinfo{year}{2015}.
\newblock \bibinfo{title}{{Mass Calibration and Cosmological Analysis of the
  SPT-SZ Galaxy Cluster Sample Using Velocity Dispersion $\sigma_v$ and X-ray
  $Y_\textrm{X}$ Measurements}}.
\newblock \bibinfo{journal}{Astrophys. J.} \bibinfo{volume}{799},
  \bibinfo{pages}{214}.
\newblock \DOIprefix\doi{10.1088/0004-637X/799/2/214},
  \href{http://arxiv.org/abs/1407.2942}{{\tt arXiv:1407.2942}}.
\bibitem[{Bolejko et~al.(2013)Bolejko, Clarkson, Maartens, Bacon, Meures and
  Beynon}]{Bolejko:2012uj}
\bibinfo{author}{Bolejko, K.}, \bibinfo{author}{Clarkson, C.},
  \bibinfo{author}{Maartens, R.}, \bibinfo{author}{Bacon, D.},
  \bibinfo{author}{Meures, N.}, \bibinfo{author}{Beynon, E.},
  \bibinfo{year}{2013}.
\newblock \bibinfo{title}{{Antilensing: The Bright Side of Voids}}.
\newblock \bibinfo{journal}{Phys. Rev. Lett.} \bibinfo{volume}{110},
  \bibinfo{pages}{021302}.
\newblock \DOIprefix\doi{10.1103/PhysRevLett.110.021302},
  \href{http://arxiv.org/abs/1209.3142}{{\tt arXiv:1209.3142}}.
\bibitem[{Bonvin et~al.(2006)Bonvin, Durrer and Gasparini}]{Bonvin:2005ps}
\bibinfo{author}{Bonvin, C.}, \bibinfo{author}{Durrer, R.},
  \bibinfo{author}{Gasparini, M.A.}, \bibinfo{year}{2006}.
\newblock \bibinfo{title}{{Fluctuations of the luminosity distance}}.
\newblock \bibinfo{journal}{Phys. Rev.} \bibinfo{volume}{D73},
  \bibinfo{pages}{023523}.
\newblock \DOIprefix\doi{10.1103/PhysRevD.85.029901,
  10.1103/PhysRevD.73.023523},
  \href{http://arxiv.org/abs/astro-ph/0511183}{{\tt arXiv:astro-ph/0511183}}.
  \bibinfo{note}{[Erratum: Phys. Rev.D85,029901(2012)]}.
\bibitem[{{Castro} and {Quartin}(2014)}]{Castro:2014oja}
\bibinfo{author}{{Castro}, T.}, \bibinfo{author}{{Quartin}, M.},
  \bibinfo{year}{2014}.
\newblock \bibinfo{title}{{First measurement of {$\sigma$}$_{8}$ using
  supernova magnitudes only}}.
\newblock \bibinfo{journal}{\mnras} \bibinfo{volume}{443},
  \bibinfo{pages}{L6--L10}.
\newblock \DOIprefix\doi{10.1093/mnrasl/slu071},
  \href{http://arxiv.org/abs/1403.0293}{{\tt arXiv:1403.0293}}.
\bibitem[{Challinor and van Leeuwen(2002)}]{Challinor:2002zh}
\bibinfo{author}{Challinor, A.}, \bibinfo{author}{van Leeuwen, F.},
  \bibinfo{year}{2002}.
\newblock \bibinfo{title}{{Peculiar velocity effects in high resolution
  microwave background experiments}}.
\newblock \bibinfo{journal}{Phys.Rev.} \bibinfo{volume}{D65},
  \bibinfo{pages}{103001}.
\newblock \DOIprefix\doi{10.1103/PhysRevD.65.103001},
  \href{http://arxiv.org/abs/astro-ph/0112457}{{\tt arXiv:astro-ph/0112457}}.
\bibitem[{{Conley} et~al.(2011)}]{conley2011}
\bibinfo{author}{{Conley}, A.}, et~al., \bibinfo{year}{2011}.
\newblock \bibinfo{title}{{Supernova Constraints and Systematic Uncertainties
  from the First Three Years of the Supernova Legacy Survey}}.
\newblock \bibinfo{journal}{\apjs} \bibinfo{volume}{192}, \bibinfo{pages}{1}.
\newblock \DOIprefix\doi{10.1088/0067-0049/192/1/1},
  \href{http://arxiv.org/abs/1104.1443}{{\tt arXiv:1104.1443}}.
\bibitem[{Courtin et~al.(2011)Courtin, Rasera, Alimi, Corasaniti, Boucher
  et~al.}]{Courtin:2010gx}
\bibinfo{author}{Courtin, J.}, \bibinfo{author}{Rasera, Y.},
  \bibinfo{author}{Alimi, J.M.}, \bibinfo{author}{Corasaniti, P.S.},
  \bibinfo{author}{Boucher, V.}, et~al., \bibinfo{year}{2011}.
\newblock \bibinfo{title}{{Imprints of dark energy on cosmic structure
  formation: II) Non-Universality of the halo mass function}}.
\newblock \bibinfo{journal}{Mon.Not.Roy.Astron.Soc.} \bibinfo{volume}{410},
  \bibinfo{pages}{1911--1931}.
\newblock \DOIprefix\doi{10.1111/j.1365-2966.2010.17573.x},
  \href{http://arxiv.org/abs/1001.3425}{{\tt arXiv:1001.3425}}.
\bibitem[{{Davis} et~al.(2011)}]{davis2011}
\bibinfo{author}{{Davis}, T.M.}, et~al., \bibinfo{year}{2011}.
\newblock \bibinfo{title}{{The Effect of Peculiar Velocities on Supernova
  Cosmology}}.
\newblock \bibinfo{journal}{\apj} \bibinfo{volume}{741}, \bibinfo{pages}{67}.
\newblock \DOIprefix\doi{10.1088/0004-637X/741/1/67},
  \href{http://arxiv.org/abs/1012.2912}{{\tt arXiv:1012.2912}}.
\bibitem[{Dodelson and Vallinotto(2006)}]{Dodelson:2005zt}
\bibinfo{author}{Dodelson, S.}, \bibinfo{author}{Vallinotto, A.},
  \bibinfo{year}{2006}.
\newblock \bibinfo{title}{{Learning from the scatter in type ia supernovae}}.
\newblock \bibinfo{journal}{Phys.Rev.} \bibinfo{volume}{D74},
  \bibinfo{pages}{063515}.
\newblock \DOIprefix\doi{10.1103/PhysRevD.74.063515},
  \href{http://arxiv.org/abs/astro-ph/0511086}{{\tt arXiv:astro-ph/0511086}}.
\bibitem[{{Eisenstein} et~al.(2005)}]{eisenstein2005}
\bibinfo{author}{{Eisenstein}, D.J.}, et~al., \bibinfo{year}{2005}.
\newblock \bibinfo{title}{{Detection of the Baryon Acoustic Peak in the
  Large-Scale Correlation Function of SDSS Luminous Red Galaxies}}.
\newblock \bibinfo{journal}{\apj} \bibinfo{volume}{633},
  \bibinfo{pages}{560--574}.
\newblock \DOIprefix\doi{10.1086/466512},
  \href{http://arxiv.org/abs/arXiv:astro-ph/0501171}{{\tt
  arXiv:arXiv:astro-ph/0501171}}.
\bibitem[{{Gordon} et~al.(2007){Gordon}, {Land} and {Slosar}}]{gordon2007}
\bibinfo{author}{{Gordon}, C.}, \bibinfo{author}{{Land}, K.},
  \bibinfo{author}{{Slosar}, A.}, \bibinfo{year}{2007}.
\newblock \bibinfo{title}{{Cosmological Constraints from Type Ia Supernovae
  Peculiar Velocity Measurements}}.
\newblock \bibinfo{journal}{Physical Review Letters} \bibinfo{volume}{99},
  \bibinfo{pages}{081301}.
\newblock \DOIprefix\doi{10.1103/PhysRevLett.99.081301},
  \href{http://arxiv.org/abs/0705.1718}{{\tt arXiv:0705.1718}}.
\bibitem[{Guy et~al.(2007)Guy, Astier, Baumont, Hardin, Pain
  et~al.}]{Guy:2007dv}
\bibinfo{author}{Guy, J.}, \bibinfo{author}{Astier, P.},
  \bibinfo{author}{Baumont, S.}, \bibinfo{author}{Hardin, D.},
  \bibinfo{author}{Pain, R.}, et~al., \bibinfo{year}{2007}.
\newblock \bibinfo{title}{{SALT2: Using distant supernovae to improve the use
  of Type Ia supernovae as distance indicators}}.
\newblock \bibinfo{journal}{Astron.Astrophys.} \bibinfo{volume}{466},
  \bibinfo{pages}{11--21}.
\newblock \DOIprefix\doi{10.1051/0004-6361:20066930},
  \href{http://arxiv.org/abs/astro-ph/0701828}{{\tt arXiv:astro-ph/0701828}}.
\bibitem[{{Hadjiyska} et~al.(2012){Hadjiyska}, {Rabinowitz}, {Baltay},
  {Ellman}, {Nugent}, {Zinn}, {Horowitz}, {McKinnon} and
  {Miller}}]{2012IAUS..285..324H}
\bibinfo{author}{{Hadjiyska}, E.}, \bibinfo{author}{{Rabinowitz}, D.},
  \bibinfo{author}{{Baltay}, C.}, \bibinfo{author}{{Ellman}, N.},
  \bibinfo{author}{{Nugent}, P.}, \bibinfo{author}{{Zinn}, R.},
  \bibinfo{author}{{Horowitz}, B.}, \bibinfo{author}{{McKinnon}, R.},
  \bibinfo{author}{{Miller}, L.R.}, \bibinfo{year}{2012}.
\newblock \bibinfo{title}{{La Silla-QUEST Variability Survey in the Southern
  Hemisphere}}, in: \bibinfo{editor}{{Griffin}, E.},
  \bibinfo{editor}{{Hanisch}, R.}, \bibinfo{editor}{{Seaman}, R.} (Eds.),
  \bibinfo{booktitle}{New Horizons in Time Domain Astronomy}, pp.
  \bibinfo{pages}{324--326}.
\newblock \DOIprefix\doi{10.1017/S1743921312000944},
  \href{http://arxiv.org/abs/1210.1584}{{\tt arXiv:1210.1584}}.
\bibitem[{Hamana and Futamase(2000)}]{Hamana:1999rk}
\bibinfo{author}{Hamana, T.}, \bibinfo{author}{Futamase, T.},
  \bibinfo{year}{2000}.
\newblock \bibinfo{title}{{A new measure of $sigma_8$ using the lensing
  dispersion in high-z type Ia sne}}.
\newblock \bibinfo{journal}{ApJ} \bibinfo{volume}{534},
  \bibinfo{pages}{29--33}.
\newblock \DOIprefix\doi{10.1086/308758},
  \href{http://arxiv.org/abs/astro-ph/9912319}{{\tt arXiv:astro-ph/9912319}}.
\bibitem[{{Hamuy} et~al.(1996){Hamuy}, {Phillips}, {Suntzeff}, {Schommer},
  {Maza} and {Aviles}}]{hamuy1996}
\bibinfo{author}{{Hamuy}, M.}, \bibinfo{author}{{Phillips}, M.M.},
  \bibinfo{author}{{Suntzeff}, N.B.}, \bibinfo{author}{{Schommer}, R.A.},
  \bibinfo{author}{{Maza}, J.}, \bibinfo{author}{{Aviles}, R.},
  \bibinfo{year}{1996}.
\newblock \bibinfo{title}{The absolute luminosities of the calan/tololo type
  {Ia} supernovae}.
\newblock \bibinfo{journal}{\aj} \bibinfo{volume}{112},
  \bibinfo{pages}{2391--2397}.
\newblock \DOIprefix\doi{10.1086/118190},
  \href{http://arxiv.org/abs/arXiv:astro-ph/9609059}{{\tt
  arXiv:arXiv:astro-ph/9609059}}.
\bibitem[{Hamuy et~al.(2006)}]{Hamuy:2005tf}
\bibinfo{author}{Hamuy, M.}, et~al., \bibinfo{year}{2006}.
\newblock \bibinfo{title}{{The carnegie supernova project: the low-redshift
  survey}}.
\newblock \bibinfo{journal}{Publ. Astron. Soc. Pac.} \bibinfo{volume}{118},
  \bibinfo{pages}{2--20}.
\newblock \DOIprefix\doi{10.1086/500228},
  \href{http://arxiv.org/abs/astro-ph/0512039}{{\tt arXiv:astro-ph/0512039}}.
\bibitem[{Hannestad et~al.(2008)Hannestad, Haugboelle and
  Thomsen}]{Hannestad:2007fb}
\bibinfo{author}{Hannestad, S.}, \bibinfo{author}{Haugboelle, T.},
  \bibinfo{author}{Thomsen, B.}, \bibinfo{year}{2008}.
\newblock \bibinfo{title}{{Precision measurements of large scale structure with
  future type Ia supernova surveys}}.
\newblock \bibinfo{journal}{JCAP} \bibinfo{volume}{0802}, \bibinfo{pages}{022}.
\newblock \DOIprefix\doi{10.1088/1475-7516/2008/02/022},
  \href{http://arxiv.org/abs/0705.0979}{{\tt arXiv:0705.0979}}.
\bibitem[{Hilbert et~al.(2008)Hilbert, White, Hartlap and
  Schneider}]{Hilbert:2007jd}
\bibinfo{author}{Hilbert, S.}, \bibinfo{author}{White, S.D.},
  \bibinfo{author}{Hartlap, J.}, \bibinfo{author}{Schneider, P.},
  \bibinfo{year}{2008}.
\newblock \bibinfo{title}{{Strong lensing optical depths in a LCDM Universe. 2.
  The influence of the stellar mass in galaxies}}.
\newblock \bibinfo{journal}{Mon.Not.Roy.Astron.Soc.} \bibinfo{volume}{386},
  \bibinfo{pages}{1845--1854}.
\newblock \DOIprefix\doi{10.1111/j.1365-2966.2008.13190.x},
  \href{http://arxiv.org/abs/0712.1593}{{\tt arXiv:0712.1593}}.
\bibitem[{Hoffman et~al.(2015)Hoffman, Courtois and Tully}]{Hoffman:2015waa}
\bibinfo{author}{Hoffman, Y.}, \bibinfo{author}{Courtois, H.M.},
  \bibinfo{author}{Tully, R.B.}, \bibinfo{year}{2015}.
\newblock \bibinfo{title}{{Cosmic Bulk Flow and the Local Motion from
  Cosmicflows-2}}.
\newblock \bibinfo{journal}{Mon. Not. Roy. Astron. Soc.} \bibinfo{volume}{449},
  \bibinfo{pages}{4494--4505}.
\newblock \DOIprefix\doi{10.1093/mnras/stv615},
  \href{http://arxiv.org/abs/1503.05422}{{\tt arXiv:1503.05422}}.
\bibitem[{Holz and Linder(2005)}]{Holz:2004xx}
\bibinfo{author}{Holz, D.E.}, \bibinfo{author}{Linder, E.V.},
  \bibinfo{year}{2005}.
\newblock \bibinfo{title}{{Safety in numbers: Gravitational lensing degradation
  of the luminosity distance-redshift relation}}.
\newblock \bibinfo{journal}{Astrophys.J.} \bibinfo{volume}{631},
  \bibinfo{pages}{678--688}.
\newblock \DOIprefix\doi{10.1086/432085},
  \href{http://arxiv.org/abs/astro-ph/0412173}{{\tt arXiv:astro-ph/0412173}}.
\bibitem[{Hudson et~al.(2004)Hudson, Smith, Lucey and
  Branchini}]{Hudson:2004et}
\bibinfo{author}{Hudson, M.J.}, \bibinfo{author}{Smith, R.J.},
  \bibinfo{author}{Lucey, J.R.}, \bibinfo{author}{Branchini, E.},
  \bibinfo{year}{2004}.
\newblock \bibinfo{title}{{Streaming motions of galaxy clusters within 12000 km
  s**-1. 5. The Peculiar velocity field}}.
\newblock \bibinfo{journal}{Mon. Not. Roy. Astron. Soc.} \bibinfo{volume}{352},
  \bibinfo{pages}{61}.
\newblock \DOIprefix\doi{10.1111/j.1365-2966.2004.07893.x},
  \href{http://arxiv.org/abs/astro-ph/0404386}{{\tt arXiv:astro-ph/0404386}}.
\bibitem[{{Hui} and {Greene}(2006)}]{hui2006}
\bibinfo{author}{{Hui}, L.}, \bibinfo{author}{{Greene}, P.B.},
  \bibinfo{year}{2006}.
\newblock \bibinfo{title}{{Correlated fluctuations in luminosity distance and
  the importance of peculiar motion in supernova surveys}}.
\newblock \bibinfo{journal}{\prd} \bibinfo{volume}{73},
  \bibinfo{pages}{123526}.
\newblock \DOIprefix\doi{10.1103/PhysRevD.73.123526},
  \href{http://arxiv.org/abs/astro-ph/0512159}{{\tt arXiv:astro-ph/0512159}}.
\bibitem[{Huterer et~al.(2015)Huterer, Shafer and Schmidt}]{Huterer:2015gpa}
\bibinfo{author}{Huterer, D.}, \bibinfo{author}{Shafer, D.L.},
  \bibinfo{author}{Schmidt, F.}, \bibinfo{year}{2015}.
\newblock \bibinfo{title}{{No evidence for bulk velocity from type Ia
  supernovae}} \href{http://arxiv.org/abs/1509.04708}{{\tt arXiv:1509.04708}}.
\bibitem[{Iocco et~al.(2009)Iocco, Mangano, Miele, Pisanti and
  Serpico}]{Iocco:2008va}
\bibinfo{author}{Iocco, F.}, \bibinfo{author}{Mangano, G.},
  \bibinfo{author}{Miele, G.}, \bibinfo{author}{Pisanti, O.},
  \bibinfo{author}{Serpico, P.D.}, \bibinfo{year}{2009}.
\newblock \bibinfo{title}{{Primordial Nucleosynthesis: from precision cosmology
  to fundamental physics}}.
\newblock \bibinfo{journal}{Phys. Rept.} \bibinfo{volume}{472},
  \bibinfo{pages}{1--76}.
\newblock \DOIprefix\doi{10.1016/j.physrep.2009.02.002},
  \href{http://arxiv.org/abs/0809.0631}{{\tt arXiv:0809.0631}}.
\bibitem[{Jenkins et~al.(2001)Jenkins, Frenk, White, Colberg, Cole
  et~al.}]{Jenkins:2000bv}
\bibinfo{author}{Jenkins, A.}, \bibinfo{author}{Frenk, C.},
  \bibinfo{author}{White, S.D.}, \bibinfo{author}{Colberg, J.},
  \bibinfo{author}{Cole, S.}, et~al., \bibinfo{year}{2001}.
\newblock \bibinfo{title}{{The Mass function of dark matter halos}}.
\newblock \bibinfo{journal}{Mon.Not.Roy.Astron.Soc.} \bibinfo{volume}{321},
  \bibinfo{pages}{372}.
\newblock \DOIprefix\doi{10.1046/j.1365-8711.2001.04029.x},
  \href{http://arxiv.org/abs/astro-ph/0005260}{{\tt arXiv:astro-ph/0005260}}.
\bibitem[{{J{\"o}nsson} et~al.(2010a){J{\"o}nsson}, {Dahl{\'e}n}, {Hook},
  {Goobar} and {M{\"o}rtsell}}]{Jonsson:2009jp}
\bibinfo{author}{{J{\"o}nsson}, J.}, \bibinfo{author}{{Dahl{\'e}n}, T.},
  \bibinfo{author}{{Hook}, I.}, \bibinfo{author}{{Goobar}, A.},
  \bibinfo{author}{{M{\"o}rtsell}, E.}, \bibinfo{year}{2010}a.
\newblock \bibinfo{title}{{Weighing dark matter haloes with gravitationally
  lensed supernovae}}.
\newblock \bibinfo{journal}{Mon.Not.Roy.Astron.Soc.} \bibinfo{volume}{402},
  \bibinfo{pages}{526--536}.
\newblock \DOIprefix\doi{10.1111/j.1365-2966.2009.15907.x},
  \href{http://arxiv.org/abs/0910.4098}{{\tt arXiv:0910.4098}}.
\bibitem[{{J{\"o}nsson} et~al.(2010b){J{\"o}nsson}, Sullivan, Hook, Basa,
  Carlberg et~al.}]{Jonsson:2010wx}
\bibinfo{author}{{J{\"o}nsson}, J.}, \bibinfo{author}{Sullivan, M.},
  \bibinfo{author}{Hook, I.}, \bibinfo{author}{Basa, S.},
  \bibinfo{author}{Carlberg, R.}, et~al., \bibinfo{year}{2010}b.
\newblock \bibinfo{title}{{Constraining dark matter halo properties using
  lensed SNLS supernovae}}.
\newblock \bibinfo{journal}{Mon.Not.Roy.Astron.Soc.} \bibinfo{volume}{405},
  \bibinfo{pages}{535--544}.
\newblock \DOIprefix\doi{10.1111/j.1365-2966.2010.16467.x},
  \href{http://arxiv.org/abs/1002.1374}{{\tt arXiv:1002.1374}}.
\bibitem[{Kainulainen and Marra(2009)}]{Kainulainen:2009dw}
\bibinfo{author}{Kainulainen, K.}, \bibinfo{author}{Marra, V.},
  \bibinfo{year}{2009}.
\newblock \bibinfo{title}{{A new stochastic approach to cumulative weak
  lensing}}.
\newblock \bibinfo{journal}{Phys.Rev.} \bibinfo{volume}{D80},
  \bibinfo{pages}{123020}.
\newblock \DOIprefix\doi{10.1103/PhysRevD.80.123020},
  \href{http://arxiv.org/abs/0909.0822}{{\tt arXiv:0909.0822}}.
\bibitem[{Kainulainen and Marra(2011)}]{Kainulainen:2010at}
\bibinfo{author}{Kainulainen, K.}, \bibinfo{author}{Marra, V.},
  \bibinfo{year}{2011}.
\newblock \bibinfo{title}{{Accurate Modeling of Weak Lensing with the sGL
  Method}}.
\newblock \bibinfo{journal}{Phys.Rev.} \bibinfo{volume}{D83},
  \bibinfo{pages}{023009}.
\newblock \DOIprefix\doi{10.1103/PhysRevD.83.023009},
  \href{http://arxiv.org/abs/1011.0732}{{\tt arXiv:1011.0732}}.
\bibitem[{Keller et~al.(2007)}]{Keller:2007xt}
\bibinfo{author}{Keller, S.C.}, et~al., \bibinfo{year}{2007}.
\newblock \bibinfo{title}{{SkyMapper and the Southern Sky Survey}}.
\newblock \bibinfo{journal}{Publ. Astron. Soc. Austral.} \bibinfo{volume}{24},
  \bibinfo{pages}{1--12}.
\newblock \DOIprefix\doi{10.1071/AS07001},
  \href{http://arxiv.org/abs/astro-ph/0702511}{{\tt arXiv:astro-ph/0702511}}.
\bibitem[{Kosowsky and Kahniashvili(2011)}]{Kosowsky:2010jm}
\bibinfo{author}{Kosowsky, A.}, \bibinfo{author}{Kahniashvili, T.},
  \bibinfo{year}{2011}.
\newblock \bibinfo{title}{{The Signature of Proper Motion in the Microwave
  Sky}}.
\newblock \bibinfo{journal}{Phys.Rev.Lett.} \bibinfo{volume}{106},
  \bibinfo{pages}{191301}.
\newblock \DOIprefix\doi{10.1103/PhysRevLett.106.191301},
  \href{http://arxiv.org/abs/1007.4539}{{\tt arXiv:1007.4539}}.
\bibitem[{{Lahav} et~al.(1991){Lahav}, {Lilje}, {Primack} and
  {Rees}}]{Lahav:1991}
\bibinfo{author}{{Lahav}, O.}, \bibinfo{author}{{Lilje}, P.B.},
  \bibinfo{author}{{Primack}, J.R.}, \bibinfo{author}{{Rees}, M.J.},
  \bibinfo{year}{1991}.
\newblock \bibinfo{title}{{Dynamical effects of the cosmological constant}}.
\newblock \bibinfo{journal}{\mnras} \bibinfo{volume}{251},
  \bibinfo{pages}{128--136}.
\bibitem[{Law et~al.(2009)}]{Law:2009ys}
\bibinfo{author}{Law, N.M.}, et~al., \bibinfo{year}{2009}.
\newblock \bibinfo{title}{{The Palomar Transient Factory: System Overview,
  Performance and First Results}}.
\newblock \bibinfo{journal}{Publ. Astron. Soc. Pac.} \bibinfo{volume}{121},
  \bibinfo{pages}{1395}.
\newblock \DOIprefix\doi{10.1086/648598},
  \href{http://arxiv.org/abs/0906.5350}{{\tt arXiv:0906.5350}}.
\bibitem[{Lewis et~al.(2000)Lewis, Challinor and Lasenby}]{Lewis:1999bs}
\bibinfo{author}{Lewis, A.}, \bibinfo{author}{Challinor, A.},
  \bibinfo{author}{Lasenby, A.}, \bibinfo{year}{2000}.
\newblock \bibinfo{title}{{Efficient Computation of CMB anisotropies in closed
  FRW models}}.
\newblock \bibinfo{journal}{Astrophys. J.} \bibinfo{volume}{538},
  \bibinfo{pages}{473--476}.
\newblock \DOIprefix\doi{10.1086/309179},
  \href{http://arxiv.org/abs/astro-ph/9911177}{{\tt arXiv:astro-ph/9911177}}.
\bibitem[{Mantz et~al.(2015)}]{Mantz:2014paa}
\bibinfo{author}{Mantz, A.B.}, et~al., \bibinfo{year}{2015}.
\newblock \bibinfo{title}{{Weighing the giants – IV. Cosmology and neutrino
  mass}}.
\newblock \bibinfo{journal}{Mon. Not. Roy. Astron. Soc.} \bibinfo{volume}{446},
  \bibinfo{pages}{2205--2225}.
\newblock \DOIprefix\doi{10.1093/mnras/stu2096},
  \href{http://arxiv.org/abs/1407.4516}{{\tt arXiv:1407.4516}}.
\bibitem[{Marra et~al.(2013)Marra, Quartin and Amendola}]{Amendola:2013twa}
\bibinfo{author}{Marra, V.}, \bibinfo{author}{Quartin, M.},
  \bibinfo{author}{Amendola, L.}, \bibinfo{year}{2013}.
\newblock \bibinfo{title}{{Accurate weak lensing of standard candles. I.
  Flexible cosmological fits}}.
\newblock \bibinfo{journal}{Phys.Rev.} \bibinfo{volume}{D88},
  \bibinfo{pages}{063004}.
\newblock \DOIprefix\doi{10.1103/PhysRevD.88.063004},
  \href{http://arxiv.org/abs/1304.7689}{{\tt arXiv:1304.7689}}.
\bibitem[{Neill et~al.(2007)Neill, Hudson and Conley}]{Neill:2007fh}
\bibinfo{author}{Neill, J.D.}, \bibinfo{author}{Hudson, M.J.},
  \bibinfo{author}{Conley, A.J.} (\bibinfo{collaboration}{SNLS}),
  \bibinfo{year}{2007}.
\newblock \bibinfo{title}{{The Peculiar Velocities of Local Type Ia Supernovae
  and their Impact on Cosmology}}.
\newblock \bibinfo{journal}{Astrophys. J.} \bibinfo{volume}{661},
  \bibinfo{pages}{L123}.
\newblock \DOIprefix\doi{10.1086/518808},
  \href{http://arxiv.org/abs/0704.1654}{{\tt arXiv:0704.1654}}.
\bibitem[{Notari and Quartin(2012)}]{Notari:2011sb}
\bibinfo{author}{Notari, A.}, \bibinfo{author}{Quartin, M.},
  \bibinfo{year}{2012}.
\newblock \bibinfo{title}{{Measuring our Peculiar Velocity by 'Pre-deboosting'
  the CMB}}.
\newblock \bibinfo{journal}{JCAP} \bibinfo{volume}{1202}, \bibinfo{pages}{026}.
\newblock \DOIprefix\doi{10.1088/1475-7516/2012/02/026},
  \href{http://arxiv.org/abs/1112.1400}{{\tt arXiv:1112.1400}}.
\bibitem[{{Perlmutter} et~al.(1999){Perlmutter}, others and {The Supernova
  Cosmology Project}}]{perlmutter1999}
\bibinfo{author}{{Perlmutter}, S.}, \bibinfo{author}{others},
  \bibinfo{author}{{The Supernova Cosmology Project}}, \bibinfo{year}{1999}.
\newblock \bibinfo{title}{Measurements of {O}mega and {L}ambda from 42
  high-redshift supernovae}.
\newblock \bibinfo{journal}{\apj} \bibinfo{volume}{517},
  \bibinfo{pages}{565--586}.
\newblock \DOIprefix\doi{10.1086/307221},
  \href{http://arxiv.org/abs/arXiv:astro-ph/9812133}{{\tt
  arXiv:arXiv:astro-ph/9812133}}.
\bibitem[{Quartin et~al.(2014)Quartin, Marra and Amendola}]{Quartin:2013moa}
\bibinfo{author}{Quartin, M.}, \bibinfo{author}{Marra, V.},
  \bibinfo{author}{Amendola, L.}, \bibinfo{year}{2014}.
\newblock \bibinfo{title}{{Accurate Weak Lensing of Standard Candles. II.
  Measuring sigma8 with Supernovae}}.
\newblock \bibinfo{journal}{Phys.Rev.} \bibinfo{volume}{D89},
  \bibinfo{pages}{023009}.
\newblock \DOIprefix\doi{10.1103/PhysRevD.89.023009},
  \href{http://arxiv.org/abs/1307.1155}{{\tt arXiv:1307.1155}}.
\bibitem[{{Riess} et~al.(1996){Riess}, {Press} and {Kirshner}}]{riess1996}
\bibinfo{author}{{Riess}, A.G.}, \bibinfo{author}{{Press}, W.H.},
  \bibinfo{author}{{Kirshner}, R.P.}, \bibinfo{year}{1996}.
\newblock \bibinfo{title}{A precise distance indicator: {T}ype {Ia} supernova
  multicolor light-curve shapes}.
\newblock \bibinfo{journal}{\apj} \bibinfo{volume}{473},
  \bibinfo{pages}{88--109}.
\newblock \DOIprefix\doi{10.1086/178129},
  \href{http://arxiv.org/abs/arXiv:astro-ph/9604143}{{\tt
  arXiv:arXiv:astro-ph/9604143}}.
\bibitem[{{Riess} et~al.(1998)}]{riess1998}
\bibinfo{author}{{Riess}, A.G.}, et~al., \bibinfo{year}{1998}.
\newblock \bibinfo{title}{Observational evidence from supernovae for an
  accelerating universe and a cosmological constant}.
\newblock \bibinfo{journal}{\aj} \bibinfo{volume}{116},
  \bibinfo{pages}{1009--1038}.
\newblock \DOIprefix\doi{10.1086/300499},
  \href{http://arxiv.org/abs/arXiv:astro-ph/9805201}{{\tt
  arXiv:arXiv:astro-ph/9805201}}.
\bibitem[{Sheth and Tormen(1999)}]{Sheth:1999mn}
\bibinfo{author}{Sheth, R.K.}, \bibinfo{author}{Tormen, G.},
  \bibinfo{year}{1999}.
\newblock \bibinfo{title}{{Large scale bias and the peak background split}}.
\newblock \bibinfo{journal}{Mon.Not.Roy.Astron.Soc.} \bibinfo{volume}{308},
  \bibinfo{pages}{119}.
\newblock \DOIprefix\doi{10.1046/j.1365-8711.1999.02692.x},
  \href{http://arxiv.org/abs/astro-ph/9901122}{{\tt arXiv:astro-ph/9901122}}.
\bibitem[{Takahashi et~al.(2011)Takahashi, Oguri, Sato and
  Hamana}]{Takahashi:2011qd}
\bibinfo{author}{Takahashi, R.}, \bibinfo{author}{Oguri, M.},
  \bibinfo{author}{Sato, M.}, \bibinfo{author}{Hamana, T.},
  \bibinfo{year}{2011}.
\newblock \bibinfo{title}{{Probability Distribution Functions of Cosmological
  Lensing: Convergence, Shear, and Magnification}}.
\newblock \bibinfo{journal}{Astrophys.J.} \bibinfo{volume}{742},
  \bibinfo{pages}{15}.
\newblock \DOIprefix\doi{10.1088/0004-637X/742/1/15},
  \href{http://arxiv.org/abs/1106.3823}{{\tt arXiv:1106.3823}}.
\bibitem[{Valageas(2000)}]{Valageas:1999ir}
\bibinfo{author}{Valageas, P.}, \bibinfo{year}{2000}.
\newblock \bibinfo{title}{{Statistical properties of the convergence due to
  weak gravitational lensing by nonlinear structures}}.
\newblock \bibinfo{journal}{Astron.Astrophys.} \bibinfo{volume}{356},
  \bibinfo{pages}{771}.
\newblock \href{http://arxiv.org/abs/astro-ph/9911336}{{\tt
  arXiv:astro-ph/9911336}}.
\bibitem[{Yoon and Huterer(2015)}]{Yoon:2015lta}
\bibinfo{author}{Yoon, M.}, \bibinfo{author}{Huterer, D.},
  \bibinfo{year}{2015}.
\newblock \bibinfo{title}{{Kinematic dipole detection with galaxy surveys:
  forecasts and requirements}}.
\newblock \bibinfo{journal}{Astrophys. J.} \bibinfo{volume}{813},
  \bibinfo{pages}{L18}.
\newblock \DOIprefix\doi{10.1088/2041-8205/813/1/L18},
  \href{http://arxiv.org/abs/1509.05374}{{\tt arXiv:1509.05374}}.
\bibitem[{Zhao et~al.(2009)Zhao, Jing, Mo and Boerner}]{Zhao:2008wd}
\bibinfo{author}{Zhao, D.}, \bibinfo{author}{Jing, Y.}, \bibinfo{author}{Mo,
  H.}, \bibinfo{author}{Boerner, G.}, \bibinfo{year}{2009}.
\newblock \bibinfo{title}{{Accurate universal models for the mass accretion
  histories and concentrations of dark matter halos}}.
\newblock \bibinfo{journal}{Astrophys.J.} \bibinfo{volume}{707},
  \bibinfo{pages}{354--369}.
\newblock \DOIprefix\doi{10.1088/0004-637X/707/1/354},
  \href{http://arxiv.org/abs/0811.0828}{{\tt arXiv:0811.0828}}.

\end{thebibliography}

\appendix

\section{MCMC details}\label{sec:MCMC}

In this paper we used a Markov Chain Monte Carlo (MCMC) method to obtain samples from the posterior distribution of the parameters. Our MCMC code consists in a flexible implementation of the Metropolis–Hastings algorithm. It has been previously tested with many different likelihood functions, confronting the results with a simpler, grid-based likelihood code.

The chains used in this work were systematically produced following a careful algorithm.  For any chain we always run a previous chain, where we discarded the first half of the points (the MCMC ``burn-in'' stage) and computed the covariance from the last points. Subsequently, we run a new chain from where we had previously stopped using the covariance of the previous chain as the covariance of our trial function. In that way our final chain was not only burn-in free but also more efficient due to the more accurate trial function.

All the chains are available for download.\footnote{\url{http://sites.if.ufrj.br/castro/en/pesquisa/artigos/}}

\section{Search for systematics}\label{sec:tests}

We tested the possibility of a few unresolved systematics being present in the JLA catalog, and whether the supernovae PV analysis demands extra parameters in order to produce better fits (or in better agreement with our mock results).

Regarding possible unresolved systematics we realize that low-z supernovae comes from both SDSS sub-sample as well as the so-called low-z sample. The latter is a compilation of several SNe observed from different surveys with different technical specification. Suspecting that this compilation could add some spurious correlation we analyzed both SDSS and low-z  SNe separately. However, the results did not present any noticeable difference.

Regarding the need for extra nuisance variables in the SN PV analysis, we tested adding a new variable $\lambda$ that parametrizes the non-linear peculiar velocities as $\sigma_{PV}(z)=\sigma_{PV,0}+\lambda z$. This could be physically motivated as non-linear velocities of deeper supernovae should be smaller since deviations of $P(k)$ from the linear approximation also becomes smaller; again, our results did not present any noticeable change in the cosmological parameters, and the constraints on $\lambda$ were fully consistent with $0$.

Finally, as discussed in Section~\ref{sec:sigma}, we also investigated whether adding back the bin with $z \in [0.9,1.0]$ changed the results significantly. We found that it did not. The most significant change was a slight shift upwards for $\sigma_8$. After marginalizing over all other parameters (including $\gamma$) we got $\sigma_8 = 0.79 \pm 0.28$.

\section{Mock Catalogs}\label{sec:mock-app}

To generate the mock catalogs, we first assumed a magnitude distribution given by a fiducial $\Lambda$CDM model in GR (i.e., $\gamma=0.55$) with $\{\sigma_8 = 0.85,\, \Omega_{m0}=0.289,\, h =0.7\}$. Then for all SNe with $z\le0.1$ we added the PV effects, employing the code made available in~\cite{hui2006}
to compute the covariance. We added to the diagonal of the covariance an intrinsic dispersion of $\sigma_{\rm int}=0.13$ mag and used the inverse of the resulting covariance to draw random realizations of a multi-normal Gaussian. For all SNe with $0.1<z\le 0.9$ we added instead uncorrelated random dispersions given by the MeMo covariance matrix of~\eqref{eq:memo-lhood}, also computed assuming $\sigma_{\rm int}=0.13$ mag (and $\,\mu_{N, {\rm int}}=0\,$ for all $N>2$).

Although the lensing PDF is sometimes approximated by a log-normal distribution, this is not a good enough approximation to accurately compute the MeMo lensing covariance when using the first 4 moments of the distribution. In particular~\cite{Amendola:2013twa} provided fits for both the central moments of the lensing PDF and for the approximate log-normal distribution. The latter, however, does not provide accurate prescriptions for the higher moments. It introduces biases in the analysis and it is unsuitable for generating good mock catalogs. For mocks, one instead has to rely on the full results for the first 8 moments computed with \texttt{turboGL} (see Section~\ref{sec:combined}). We thus turned to the results originally obtained in~\cite{Amendola:2014yca}, where \texttt{turboGL} was used to compute the lensing moments as a function of $\{z,\,\Omega_{m0},\,\sigma_8,\,\gamma\}$.

Finally, we combined the lensing simulations described above with the PV covariance using the code created by~\cite{hui2006}, as described in Section~\ref{sec:combined}.


\end{document}